\documentclass[aps,prl,twocolumn,superscriptaddress,showpacs,amsmath,amssymb, amsfonts,10pt]{revtex4-2}
\usepackage{graphicx}
\usepackage[unicode=true,bookmarks=true,bookmarksnumbered=true,bookmarksopen=false, breaklinks=false,pdfborder={0 0 0},pdfborderstyle={},backref=false,colorlinks=false]{hyperref} 
\hypersetup{
  pdftitle={Quantum Simulation of Semiconductor Excitons in Ultracold Dipolar Fermi Gases},
  pdfauthor={Florian Hirsch; Oriana K. Diessel; Rafał Ołdziejewski; Richard Schmidt},
  pdfsubject={Research article},
  pdfkeywords={excitons, ultracold atoms, dipolar fermions, quantum simulation, optical lattices}
}
\usepackage{ragged2e}
\usepackage{braket}

\newcommand{\e}[0]{\mathrm{e}}
\newcommand{\ii}[0]{\mathrm{i}}
\newcommand{\kk}[0]{{\mathbf{k}}}
\newcommand{\KK}[0]{{\mathbf{K}}}

\newcommand{\GGamma}[0]{\mathbf{\Gamma}}
\newcommand{\qq}[0]{{\mathbf{q}}}

\newcommand{\pp}[0]{{\mathbf{p}}}
\newcommand{\rr}[0]{{\mathbf{r}}}

\newcommand{\dd}[0]{{\mathbf{d}}}
\newcommand{\RR}[0]{{\mathbf{R}}}

\newcommand{\ddelta}[0]{{\boldsymbol{\delta}}}
\newcommand{\eeta}[0]{{\boldsymbol{\eta}}}

\begin{document}

\title{Quantum Simulation of Semiconductor Excitons in Ultracold Dipolar Fermi Gases}

\author{Florian Hirsch}
\email[Email: ]{hirsch@thphys.uni-heidelberg.de}
\affiliation{Institute for Theoretical Physics, Heidelberg University, Heidelberg, Germany}
\affiliation{Center for the Physical Foundations of Computation, Heidelberg University, Heidelberg, Germany}
\author{Oriana K. Diessel}
\affiliation{ITAMP, Center for Astrophysics, Harvard \& Smithsonian, Cambridge, MA 02138, USA}
\affiliation{Department of Physics, Harvard University, Cambridge, MA 02138, USA}
\author{Rafa\l{} O\l{}dziejewski}
\affiliation{Centre for Quantum Optical Technologies, Centre of New Technologies, University of Warsaw, Banacha 2c, 02-097 Warsaw, Poland}
\author{Richard Schmidt}
\affiliation{Institute for Theoretical Physics, Heidelberg University, Heidelberg, Germany}
\affiliation{Center for the Physical Foundations of Computation, Heidelberg University, Heidelberg, Germany}

\date{July 21, 2026}
\begin{abstract}
\noindent  Inspired by the progress on the study of exciton physics in atomically thin transition metal dichalcogenide (TMD) semiconductors, we investigate the formation of analogs of excitons in cold atomic systems. To this end, we consider single-component fermions comprised of ultracold ground-state molecules or dipolar atoms in a hexagonal optical lattice. An energy offset between triangular sublattices opens up a band gap with degeneracies at the K/K$^\prime$ points as in TMDs. We predict the existence of cold atomic excitons and show that cold atoms allow us to study excitons from the weak-coupling regime, where effective mass models apply, to the strong-interaction regime, described by flat-band models.
We demonstrate how these excitons can be observed using lattice modulation spectroscopy, and how their wave functions can be mapped out using quantum gas microscopy. 
Firmly establishing the  idea of quantum simulation of semiconductor physics, this work lays the foundation for simulating complex electronic states such as trions, polarons and excitonic insulators. 
\end{abstract}

\maketitle 

In the past two decades, ultracold atoms have emerged as a powerful platform for simulating condensed matter phenomena, offering insights into effects difficult to analyze in detail in solid-state systems 
\cite{jaksch1998cold,  greiner2002quantum, schafer2020tools,  yang2020cooling, altman2021quantum, su2023dipolar,  meng2023atomic, langen2024quantum, cornish2024quantum, halimeh2025cold, chalopin2025optical}.
Simultaneously, in solid-state physics, two-dimensional (2D) materials have gained attention as a novel avenue for exploring correlated states of matter with potential technological applications \cite{novoselov2004electric, geim2007rise, geim2013van, fiori2014electronics}. This applies in particular to transition metal dichalcogenides (TMDs), which combine the  physics of semiconductors and 2D materials \cite{wang2012electronics, fang2015ab, Kogar2017, cotlet2019transport, fey2020theory, Trovatello2020,imamoglu2021exciton}.
TMDs feature excitons of remarkable stability \cite{chernikov2014exciton}, enabling a wide range of optoelectronic applications \cite{wang2018colloquium, qiu2013optical, Mak2013, Courtade2017}. 
Driven by this progress, many open questions have emerged, ranging from the existence  of excitonic insulators \cite{ma2021strongly, sun2022evidence, kaneko2025new} to novel mechanisms of exciton-induced superconductivity \cite{laussy2012superconductivity, crepel2021new, von2024superconductivity,zerba2024realizing,crepel2023topological,vlasiuk2026enhancing}. 
However, despite the central role of semiconductors in solid-state physics and technological development, so far no quantum simulation of semiconductor physics has been realized in ultracold atoms. 

In this Letter, we propose a new pathway to address outstanding questions in solid-state materials by establishing quantum simulation of 2D materials with ultracold atoms. As a first step in emulating the complex physics of correlated electronic states in atomically thin semiconductors,  we demonstrate the existence of the atomic analog of excitons in a TMD-like band structure realized in optical lattices.  We discuss  experimental methods to probe exciton physics and show how  cold atoms  can be tuned to explore semiconductor physics across a variety of physical systems, ranging from the effective mass regime of TMDs to the flat-band regime realized in magic-angle graphene and twisted bilayer TMDs \cite{cao2018correlated,cao2018unconventional,yankowitz2019tuning, mak2022semiconductor, checkelsky2024flat}.

\paragraph{\textbf{Model.---}} 
The quantum simulation of 2D semiconductor excitons requires a few minimal ingredients: (a) a band structure with two bands, featuring (b) a band gap, such that the lower valence band can be completely filled with single-component fermions, (c) repulsive interactions between the fermions and (d) the ability to induce interband transitions of the fermions. Previous work has investigated excitons either using bosonic cold atoms \cite{zoubi2007excitons, zhang2013particle} or on top of a Mott state \cite{bohrdt2024spectroscopy}. Neither approach fulfills the requirements for the quantum simulation of semiconductor physics.

\begin{figure}
    \centering
    \includegraphics[width=\linewidth]{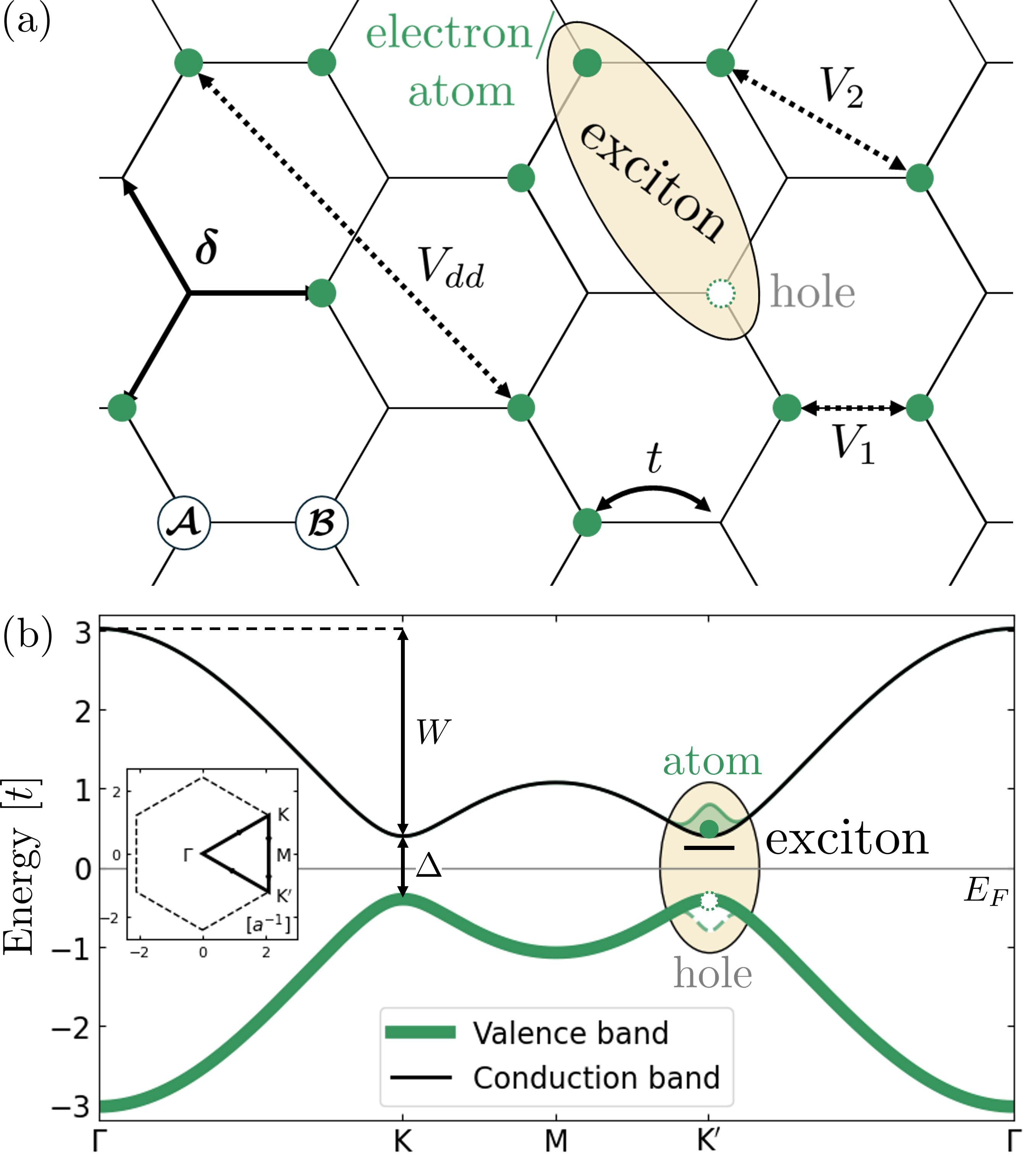}
    \caption{\justifying \textbf{Semiconductor band structure realized in cold atoms.} (a) Honeycomb lattice with two sublattices $\mathcal{A}$ and $\mathcal{B}$, half-filled with  dipolar fermions (green) hopping between sites. The fermions interact via a dipolar potential $V_\text{dd}$ and can form `electron-hole' bound exciton states. (b) Cut through the band structure on the path through the first Brillouin zone as indicated in the inset. The Fermi level resides inside the gap  and the exciton wave function  in the $\KK'$ valley is illustrated. }
    \label{fig:latticemodel}
\end{figure}

To realize a band structure similar to that of TMDs, we confine the fermions in a honeycomb lattice with energy offset $\Delta$ between the triangular sublattices $\mathcal{A}$ and $\mathcal{B}$, opening up a band gap (Fig.~\ref{fig:latticemodel}(a,b)). Experimentally, such a lattice can, e.g., be implemented using optical superlattices \cite{sebby2006lattice, aidelsburger2011experimental} or by the interference of laser beams \cite{soltan2011multi}. The resulting system can be described by a tight-binding model for the kinetic energy
\begin{equation}
    H_0 = -t \sum_{\langle i, j \rangle}  \left( a^\dagger_{i} b_{j} + b^\dagger_{j} a_{i} \right) + \frac{\Delta}{2} \sum_{i} \left( a^\dagger_{i} a_{i} - b^\dagger_{i} b_{i} \right). \label{eq:Hzerotightbinding}
\end{equation}
Here fermions can tunnel between nearest-neighbor lattice sites $i$, $j$ with hopping amplitude $t$. 
The Hamiltonian can be diagonalized, $\relpenalty=10000 H_0 = \sum_\kk \left(E_\kk^v v^\dagger_\kk v_\kk + E_\kk^c c^\dagger_\kk c_\kk\right)$, using valence and conduction band creation operators ($v^\dagger_\kk$, $c^\dagger_\kk$), respectively. Their dispersion relation $\relpenalty=10000 E^{c/v}_\kk = \pm \sqrt{\left(\frac{\Delta}{2} \right)^2 + |t D_\kk|^2}$ is depicted in Fig.~\ref{fig:latticemodel}(b), featuring a direct band gap at the K/K$^\prime$ points of the Brillouin zone; $D_\kk = \sum_{\ddelta} \e^{\ii \kk \cdot \ddelta}$ with $\ddelta$ summing over the three nearest-neighbor vectors between the $\mathcal{A}$ and $\mathcal{B}$ sites (for details see the Supplemental Material \cite{SM}\nocite{Numpy2020, Scipy2020, Matplotlib2007, combescot2015excitons, soltan2011multi, Flaeschner2018}).

Repulsive interactions are realized by  dipolar fermions (magnetic atoms or dipolar molecules) whose dipole moments $\dd$ are aligned perpendicular to the lattice using an external field.
This introduces a crucial new ingredient absent from the many alkali-atom-based studies of Fermi-Hubbard physics: the dipole-dipole repulsion generates significant beyond-onsite interactions, enabling the simulation of the physics of long-range electron-electron repulsion between equal-spin fermions that is the basis for excitons in 2D materials.

The  interaction Hamiltonian is given by
\begin{equation}
    H_{\text{int}} = \sum_{ij} \bigg[\frac{1}{2} V^{(1)}_{ij} \left(a^\dagger_{i}a^\dagger_{j}a_{j}a_{i} + b^\dagger_{i}b^\dagger_{j}b_{j}b_{i}\right) + V^{(2)}_{ij} a^\dagger_{i}b^\dagger_{j}b_{j}a_{i} \bigg] \label{eq:Hintrealspace}.
\end{equation}
Assuming Wannier orbitals well-localized on  sites  $\RR_i$, one has $V^{(1)}_{ij} = V_\text{tot}(\RR_j -\RR_i)$, $V^{(2)}_{ij} =  V_\text{tot}(\RR_j + \ddelta -\RR_i)$ with $V_\text{tot}(\rr) = V_\text{s}(\rr) + V_\text{dd}(\rr)$ where $V_\text{dd}(\rr) \sim \frac{\dd^2}{|\rr|^3}$ represents the dipolar contribution and $V_\text{s}(\rr)$ short-range details. Since the latter only become relevant at distances much smaller than the lattice spacing, they can be neglected.

Our setting explicitly does not require large filling fractions. This makes it immediately realizable with current ultracold molecule setups where achieving unit filling fractions in optical lattices remains an outstanding challenge \cite{schindewolf2022evaporation, cornish2024quantum}. Instead, our approach requires only near half-filling, and using entropy engineering, the condition $T \ll \Delta$ can be readily achieved \cite{catani2009entropy, reichsollner2017quantum, chiu2018quantum}.

\paragraph{\textbf{Exciton wave function.---}} 
We consider an optical lattice at half-filling, i.e. for $T \ll \Delta$ the valence band is filled while the conduction band is empty. We describe this state as a Slater determinant $\ket{\text{FS}}_v= \prod_\kk v^\dagger_\kk \ket{0}_v$, where $\ket{0}_v$ denotes the valence band vacuum. To model (zero-momentum) excitons, we choose the ansatz (see Fig.~\ref{fig:latticemodel}(b))
\begin{equation}
    \ket{X_n} = \sum_\pp \alpha_{n\pp} c^\dagger_\pp v_{\pp}\ket{\text{FS}}_v \ket{0}_c , \label{eq:excitonstate}
\end{equation}
describing the analog of an optical excitation of electrons from the valence to the conduction band in semiconductors. Within this restricted Hilbert space of a single particle-hole excitation, we use exact diagonalization (ED) of $\relpenalty=10000 H = H_0 + H_\text{int}$ to obtain exciton eigenenergies $E_n$ and momentum-space eigenstates $\alpha_{n\pp}$ (details are provided in the Supplemental Material~\cite{SM}).

\begin{figure}
    \centering
    \includegraphics[width=\linewidth]{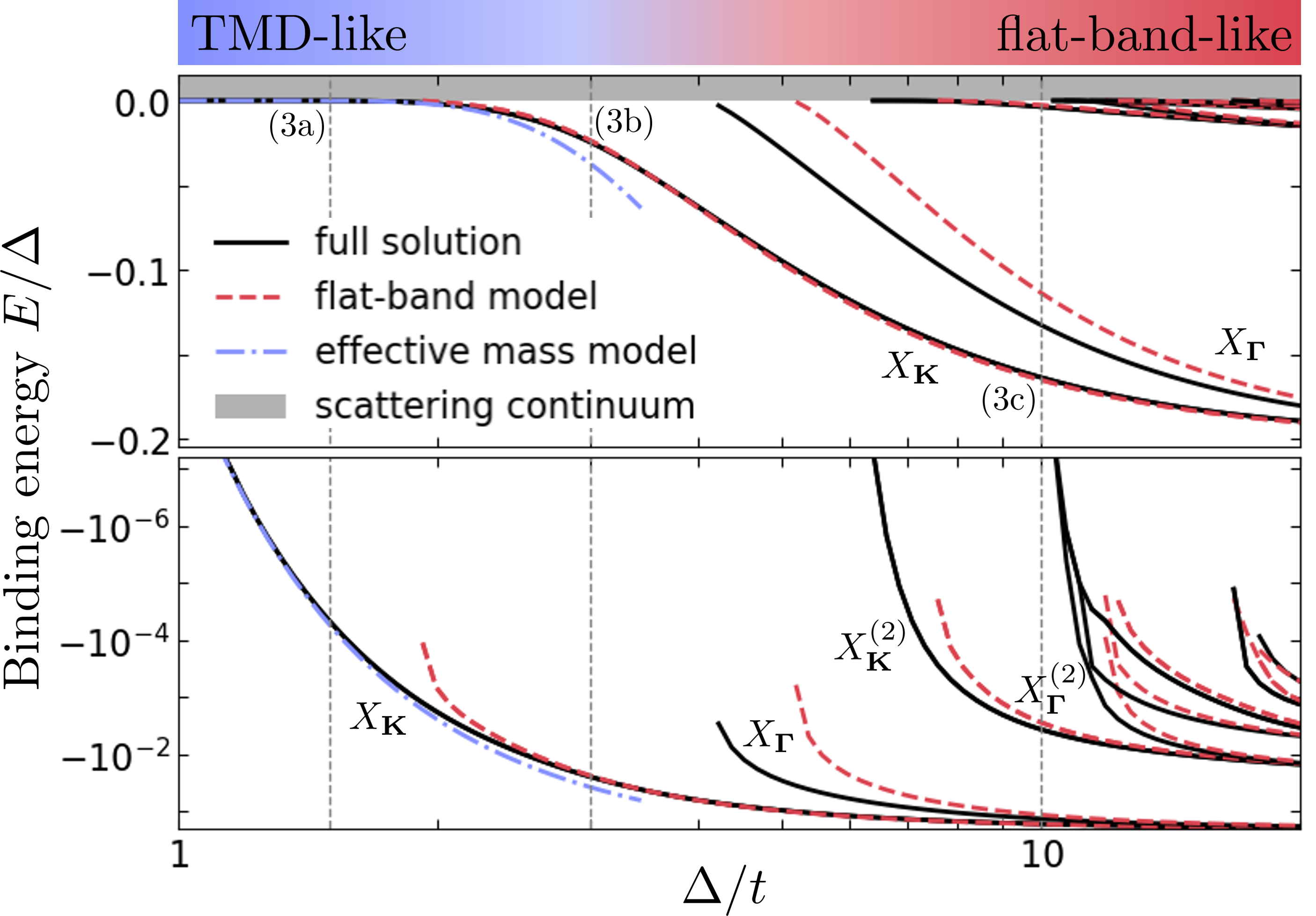}
    \caption{\justifying \textbf{Exciton binding energies.} Eigenenergies obtained from ED (black solid), compared to the flat-band model (red dashed) and effective mass model (blue, dot-dashed) as a function of $\Delta/t$, at fixed $V_1/\Delta=0.2$. Bottom: Energies on logarithmic scale. The energetically lowest exciton states are denoted as $X_\KK$ and $X_\GGamma$, followed by $X_\KK^{(2)}$ and $X_\GGamma^{(2)}$ as excited states. 
    The ground state $X_\KK$ is twofold degenerate for all $\Delta/t$. The eigenstates at the three vertical cuts are shown in Fig.~\ref{fig:boundstates}(a-c).
    } 
    \label{fig:spaghetti}
\end{figure}

\paragraph{\textbf{Exact diagonalization.---}} 
We now demonstrate the existence of excitons. Exciton binding energies obtained from ED as a function of the ratio between band gap and hopping strength $\Delta/t$ are shown in Fig.~\ref{fig:spaghetti}, which applies to both ultracold dipolar atoms and ultracold molecules. 
The tunability of the ratios $V_1/\Delta$ and $\Delta/t$, and thus the relation between kinetic and interaction energy, enables quantum simulation of a wide set of correlated electron states in one setup. 

\begin{figure}
    \centering
    \includegraphics[width=1\linewidth]{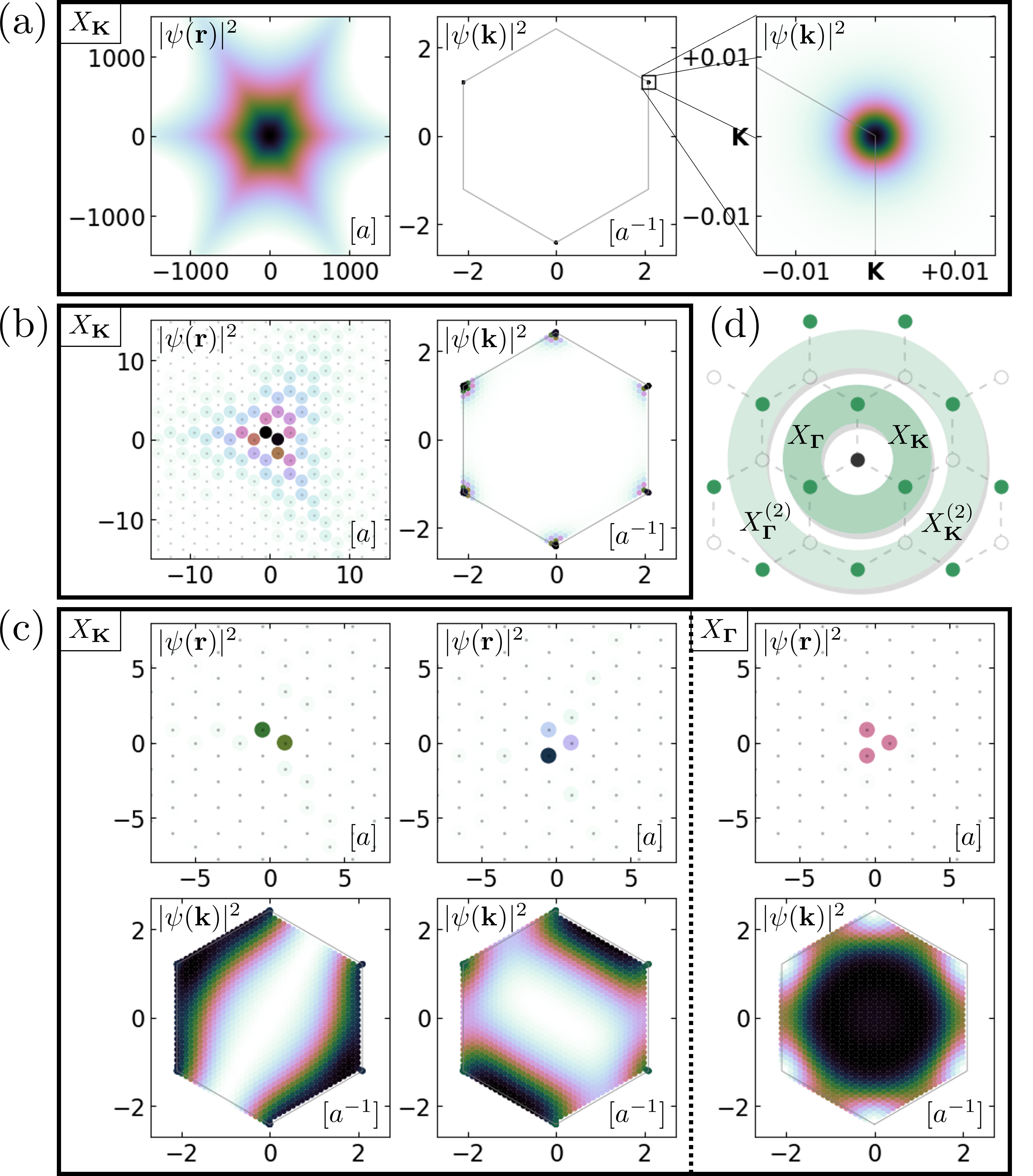}
    \caption{\justifying \textbf{Numerical results for the exciton eigenstates.} Exciton wave functions corresponding to the cuts in Fig.~\ref{fig:spaghetti}, each in real and momentum space with units $a$ and $a^{-1}$, respectively. (a) Exciton state $X_\KK$ for $\Delta/t=1.5$, including a magnification of the region around the $\KK$-point, (b) exciton state $X_\KK$ for $\Delta/t=3$, (c) exciton states $X_\KK$ (twofold degenerate) and $X_\GGamma$ for $\Delta/t=10$, (d) illustration of the emerging decoupled three-level systems in the flat-band limit.}
    \label{fig:boundstates}
\end{figure}

For small $\Delta/t$, kinetic energy dominates, and as a result exciton binding energies are small. This represents the regime realized in two-dimensional TMD semiconductors. Here the wave function indeed extends over hundreds of lattice sites, corresponding to the formation of large Wannier excitons (see Fig.~\ref{fig:boundstates}(a)): the wave function is  highly localized in momentum space and situated in the degenerate $K$, $K'$ valleys, precisely as in TMDs \cite{wang2018colloquium}. 

When hopping is small (i.e. $\Delta/t \gg 1$), one enters the flat-band regime realized in twisted 2D materials, such as twisted bilayer graphene \cite{yankowitz2019tuning} or TMD heterostructures \cite{mak2022semiconductor}. Here the wave function contracts to fewer lattice sites (Fig.~\ref{fig:boundstates}(b)), ultimately representing a Frenkel-type exciton at large values of $\Delta/t$ (see Fig.~\ref{fig:boundstates}(c)) and the momentum space wave function completely changes its structure. As discussed below, this is a direct result of  physics becoming dominated by ring-type exchange processes on single plaquettes (Fig.~\ref{fig:boundstates}(d)). Note that the parameters  for Fig.~\ref{fig:spaghetti} have been chosen to represent accessible values in current experimental setups as given in Table~\ref{tab:expparamscoldatoms}. This shows that current techniques place a wide variety of excitonic effects within experimental reach. 

To provide further intuition about the two regimes of solid-state excitons realizable in quantum simulations (i.e. effective mass and flat-band regimes), we will now derive regime-specific approximations, leading to remarkably precise asymptotic results.

\paragraph{\textbf{Effective mass approximation.---}} 
In TMDs, the effective mass approximation is an accurate model to describe excitons \cite{wang2018colloquium}. There, the bandwidth \mbox{$W = \sqrt{\left(\frac{\Delta}{2}\right)^2 + \left(3t\right)^2} - \frac{\Delta}{2}$} is much larger than the scale set by Coulomb interactions, such that only the regions close to the K/K$^\prime$ points in the Brillouin zone are energetically relevant. The dispersion in this region is approximately quadratic, leading to the simplified kinetic Hamiltonian $H_{\text{kin}}^{\text{eff}} = \sum_\pp \frac{\pp^2}{2m^*} c^\dagger_\pp c_\pp$ with $\pp = \kk - \KK$ and  $|\pp| \ll |\KK|$. In the tight-binding Hamiltonian, the effective mass $m^*$ is given by $m^* = \frac{2 \hbar^2 \Delta}{9 t^2 a^2}$ with nearest-neighbor spacing $a$ (negative sign for the valence band).

Performing a particle-hole transformation in the valence band, using $v_\kk = h^\dagger_{-\kk}$, the  filled valence band $\ket{0}_h = \ket{\text{FS}}_v$ is identified with the hole vacuum. Applying the transformation to the interaction term, and dropping all momentum-independent energy corrections, one obtains a minus sign for the interaction between valence band holes and conduction band electrons (the kinetic part of holes also acquires a minus sign). The interaction between particles and holes thus becomes an effective attraction: 
\begin{equation}
    H_{\text{int}}^{\text{eff}} = - \frac{1}{N}\sum_{\kk\kk'\qq} V_\qq c^\dagger_\kk h^\dagger_{\kk'} h_{\kk'-\qq} c_{\kk + \qq}. \label{eq:attractiveinteraction}
\end{equation}

Equation~\eqref{eq:attractiveinteraction} makes evident the intuitive picture underlying  exciton formation: the missing electron in the otherwise fully filled valence band acts like a quasiparticle hole of opposite `charge' (or in the atomic case, a `defect dipole' of effective opposite orientation), leading to attraction and the possibility of bound state formation.

The effective mass Hamiltonian with $ H_{\text{int}}^{\text{eff}}$ permits us to solve the exciton problem as a simple two-body problem in terms of a Bethe-Salpeter equation \mbox{$\left( E^c_\pp - E^v_\pp \right) \alpha_\pp - \frac{1}{N}\sum_\qq V_\qq \alpha_{\pp-\qq}  = E \alpha_\pp$}. After a Fourier transform, one arrives at the real-space Schrödinger equation for the exciton eigenstates $\psi(\rr)$,
\begin{equation}
    \left[- \frac{\hbar^2\nabla_\rr^2}{m^*} + V(\rr) \right] \psi(\rr) = E\psi(\rr),\label{eq:effectivemassreal}
\end{equation}
where $V(\rr)$ represents the dipolar potential that now has an attractive sign. Specifically, we choose $V(\rr) = -\frac{\dd^2}{|\rr|^3}$ for $|\rr|>r_0$ and $V(\rr) = -\frac{\dd^2}{r_0^3}$ for $|\rr|\le r_0$. The short-distance cutoff $r_0=0.705a$ is numerically determined by matching the numerical exciton eigenenergy at $\Delta/t=1.2$; for a detailed derivation of Eq. \eqref{eq:effectivemassreal}, see Ref.~\cite{SM}. 

Solving Eq.~\eqref{eq:effectivemassreal} yields energies that agree very well with the full solution in the range of  $\Delta/t \le 2$. Furthermore, we find our assumption of only momenta near $\KK$ contributing to be valid, and observe the real-space extent of the exciton to be very large, corresponding to the Wannier excitons found in TMDs (see Fig.~\ref{fig:boundstates}(a)).  

\begin{table}[t]
    \centering
    \begin{tabular}{l|l|l|l}
\hline
Parameter             & Erbium \cite{su2023dipolar}                    & \multicolumn{2}{c}{NaK \cite{chen2024ultracold, schindewolf2022evaporation}}         \\ \hline
Band gap $\Delta$      & up to $10$ kHz               & \multicolumn{2}{c}{up to $10$ kHz} \\
Hopping strength $t$           & up to $300$ Hz              & \multicolumn{2}{c}{up to $300$ Hz} \\
Dipole moment $\dd$ & 7 $\mu_B$ ($\sim0.1$ Debye)     & \multicolumn{2}{c}{2.72 Debye}  \\
NN distance $a$       & 266\,nm                    & 532\,nm         & 752\,nm         \\
NN interaction $V_{1}$ & 30\,Hz                   & 1\,kHz          & 600\,Hz         \\
NNN interaction $V_{2}$ & 6\,Hz                   & 200\,Hz         & 120\,Hz         \\
\hline
\end{tabular}
    \caption{\justifying Overview of experimental parameters realized in current experimental setups using dipolar atoms and molecules. For the definitions of $V_1$ and $V_2$, see Fig.~\ref{fig:latticemodel}(a).}
    \label{tab:expparamscoldatoms}
\end{table}

\paragraph{\textbf{Flat-band approximation.---}} 
For large $\Delta / t$, the bandwidth $W \sim \frac{9t^2}{\Delta}$ becomes small, again allowing for a simplified effective model. Treating the hopping term as a perturbation, we decouple subspaces with different particle numbers in the two sublattices $\mathcal{A}$ and $\mathcal{B}$ using a Schrieffer-Wolff transformation (see Ref.~\cite{SM}). After a further  particle-hole transformation, we arrive at a two-particle lattice problem, where the conduction-band electron is confined to $\mathcal{A}$ and the valence-band hole to $\mathcal{B}$ sites. As all terms in the resulting Hamiltonian depend only on the electron-hole separation, the problem reduces to a single-particle model on the Hilbert space specified by the lattice of relative positions between $\mathcal{A}$ and $\mathcal{B}$ sites (with its nearest-neighbor vectors $\eeta$),
\begin{equation}
    H_\text{FB} = \Delta+\frac{6t^2}{\Delta} + \frac{2t^2}{\Delta} \sum_i \sum_\eeta d^\dagger_i d_{i+\eeta} - \sum_i V_{\rr_i} d^\dagger_i d_{i} \,,\label{eq:hrelativeposition}
\end{equation}
with electron-hole pair creation operator $d^\dagger_i$ and exciton state $\ket{X} = \sum_i \beta_i d^\dagger_i \ket{0}_d$.

The effective model Eq.~\eqref{eq:hrelativeposition} can again be solved via numerical eigenvalue decomposition. The energies are shown as red dashed curves in Fig.~\ref{fig:spaghetti} and are in very good agreement with the full solution.

Remarkably, the model not only captures the ground state, but also the excited states in the problem.
This allows to develop an analytical understanding of the structure of these excited states: in a honeycomb lattice with the electron in sublattice $\mathcal{A}$ and the hole in sublattice $\mathcal{B}$, there are three relative positions $\rr_i$ of equal and minimal length, i.e. nearest-neighbor distance $a$. When $\frac{2t^2}{V_1\Delta}$ becomes small enough, the Hamiltonian \eqref{eq:hrelativeposition} is dominated by the potential term, which, due to the strong inverse scaling ($V_\rr\sim\frac{1}{r^3}$), energetically decouples the three minimal-distance grid points from all others (see Fig.~\ref{fig:boundstates}(d)). 

This energetic constraint leads to an emerging three-level system described by the Hamiltonian $\relpenalty=10000  H_3 = \frac{2t^2}{\Delta} \begin{pmatrix} 0 & 1 & 1 \\ 1 & 0 & 1 \\ 1 & 1 & 0 \end{pmatrix}$ up to a constant. Its eigenvalue spectrum $\frac{2t^2}{\Delta}\{-1,-1,2\}$ with a degenerate ground state explains the observed splitting of states. As can be seen from Fig.~\ref{fig:boundstates}(c), the degenerate ground state now has p-like characteristics, with $C_3$ angular momentum $m=\pm1$. This is directly explained by the model $H_\text{FB}$, as the hopping term in Eq.~\eqref{eq:hrelativeposition} now carries a \textit{positive} sign, leading to effective kinetic frustration.
The  splitting into a degenerate doublet $X_\KK$ and a single state $X_\GGamma$ maintains the K/K$^\prime$ degeneracy seen at smaller $\Delta/t$, with $X_\GGamma$ appearing as the extra state completing the three-level system. In the limit of large $\Delta/t$, the descriptions of $X_\GGamma$ from ED, the flat-band approximation Eq.~\eqref{eq:hrelativeposition} and the three-level system $H_3$ all agree. Interestingly, for $\Delta/t\approx6\text{ to }11$, a second three-level system ($X_\KK^{(2)}$, $X_\GGamma^{(2)}$) emerges from the three next-to-minimal-distance grid points with distance $2a$ and interaction strength $\frac{V_1}{8}$, once again captured well by the flat-band approximation.

\paragraph{\textbf{Exciton spectroscopy and microscopy.---}} 
Excitons in semiconductors are typically created by laser excitation inducing ‘vertical’ transitions across the band gap. This principle can be directly translated to cold atom experiments by choosing a time-periodic modulation of the optical lattice potential that closely mimics the laser-induced optical transitions in semiconductors, i.e. no momentum is transferred. Experimentally, this can be achieved by modulating the lattice-beam intensities. 

In the tight-binding formalism, this perturbation can be expressed as $\hat{O} = \sum_\qq T_\qq c^\dagger_\qq v_\qq$, with $T_\qq$ given in Ref.~\cite{SM}. Within linear response theory, the transition rate to a single particle-hole excitation is then given by Fermi's golden rule
\begin{equation}
    A(\omega) = \frac{2 \pi}{\hbar} \sum_n \left| \bra{X_n}\hat{O} \ket{\text{FS}} \right|^2 \delta(E_n - E_0 - \hbar \omega). \label{eq:fermisgoldenrule}
\end{equation}
In Ref.~\cite{SM} we show representative spectra for all regimes of Fig.~\ref{fig:boundstates}, demonstrating the ability to simulate typical optical experiments performed in semiconductor physics. 
Beyond simulating the optical response of 2D materials, the wave function of excitons can be studied in momentum space using time-of-flight imaging. The real-space wave function can in turn be mapped out using quantum gas microscopy, which, in terms of spatial resolution, goes beyond the state of the art of solid-state techniques \cite{bakr2009quantum, cheuk2015quantum, gross2021quantum}.

\paragraph{\textbf{Conclusion.---}} We have shown that cold atoms hold the potential to simulate optoelectronic semiconductor physics. 
Specifically, we have demonstrated the existence of cold atomic excitons for dipolar interactions over a large range of hopping strength and band gap energies. This opens the door towards a plethora of research questions concerning excitons that are currently of emerging interest in solid-state experiments. 
This includes the study of doping dependence, the modeling of exciton-electron or exciton-exciton interactions, as well as the analysis of temperature-dependent electron transport modified by excitons, and exciton condensation. Moreover, quantum simulations of exciton-induced binding mechanisms of electrons could provide new insight into the emergence of superconductivity from micro- to macroscopic scales.

{\textit{Acknowledgements.---}} We thank Atac Imamoglu, Xin-Yu Luo and Andrea Bergschneider for helpful discussions. We acknowledge funding by the DFG (German Research Foundation) – Project-ID 273811115 – SFB 1225 ISOQUANT, and under Germany's Excellence Strategy EXC 2181/1 - 390900948 (the Heidelberg STRUCTURES Excellence Cluster). 
F.H. acknowledges support from the Studienstiftung des deutschen Volkes. 
O.K.D. acknowledges support from the NSF through a grant for ITAMP at Harvard University. 
R.O. was supported by the ``Quantum Optical Technologies'' (FENG.02.01-IP.05-0017/23) project which is being carried out within the Measure 2.1 International Research Agendas programme of the Foundation for Polish Science co-financed by the European Union under the European Funds for Smart Economy 2021-2027 (FENG).

\end{document}


\title{Supplemental Material for \\ `Quantum Simulation of Semiconductor Excitons in Ultracold Dipolar Fermi Gases'}

\author{Florian Hirsch}
\email[Email: ]{hirsch@thphys.uni-heidelberg.de}
\affiliation{Institute for Theoretical Physics, Heidelberg University, Heidelberg, Germany}
\affiliation{Center for the Physical Foundations of Computation, Heidelberg University, Heidelberg, Germany}
\author{Oriana K. Diessel}
\affiliation{ITAMP, Center for Astrophysics, Harvard \& Smithsonian, Cambridge, MA 02138, USA}
\affiliation{Department of Physics, Harvard University, Cambridge, MA 02138, USA}
\author{Rafa\l{} O\l{}dziejewski}
\affiliation{Centre for Quantum Optical Technologies, Centre of New Technologies, University of Warsaw, Banacha 2c, 02-097 Warsaw, Poland}
\author{Richard Schmidt}
\affiliation{Institute for Theoretical Physics, Heidelberg University, Heidelberg, Germany}
\affiliation{Center for the Physical Foundations of Computation, Heidelberg University, Heidelberg, Germany}

\maketitle

In this Supplemental Material, we present detailed calculations leading to the main results and conclusions in the main text. Sec.~\ref{sec:conventions} outlines useful conventions and provides the definition of the studied lattice structure. In Sec.~\ref{sec:numericalsolution}, we present the analytical calculations underlying the numerical solution of the exciton problem. Secs.~\ref{sec:effective} and \ref{sec:flat} contain calculations for the effective theories employed in the effective mass and flat-band limit, respectively. 
Lastly, in Sec.~\ref{sec:fermi}, we derive the perturbation operator used in the calculation of the spectra, while Sec.~\ref{sec:spectra} presents numerical results for the absorption spectra for the exciton states shown in the main text.

\section{Conventions, definition of lattice structure and single-particle band structure} \label{sec:conventions}
Here we give our Fourier transform convention, define the lattice geometry and calculate its single-particle band structure.
We set $\hbar=1$ throughout this Supplemental Material.

\subsection{Fourier transform convention}
The fermionic creation operator $c^\dag_j$ at lattice site $\RR_j$ is related to its momentum-space representation $c^\dag_\kk$, where $\kk$ denotes the crystal momentum, as follows:
\begin{equation}
    \begin{aligned}
        c^\dag_\kk &= \frac{1}{\sqrt{N}} \sum_j \e^{\ii \kk \cdot \RR_j} c_j^\dag \\
        c_\kk &= \frac{1}{\sqrt{N}} \sum_j \e^{-\ii \kk \cdot \RR_j} c_j \\
        c^\dag_j &= \frac{1}{\sqrt{N}} \sum_\kk \e^{-\ii \kk \cdot \RR_j} c_\kk^\dag \\
        c_j &= \frac{1}{\sqrt{N}} \sum_\kk \e^{\ii \kk \cdot \RR_j} c_\kk,
    \end{aligned}
    \label{eq:fourierconvention}
\end{equation}
where $N$ is the number of unit cells and hence the number of sites per sublattice; the full honeycomb lattice has $2N$ sites.

\subsection{Lattice definition}
\begin{figure}[ht]
  \centering
  \begin{subfigure}[b]{0.45\linewidth}
    \centering
    \includegraphics[width=\textwidth]{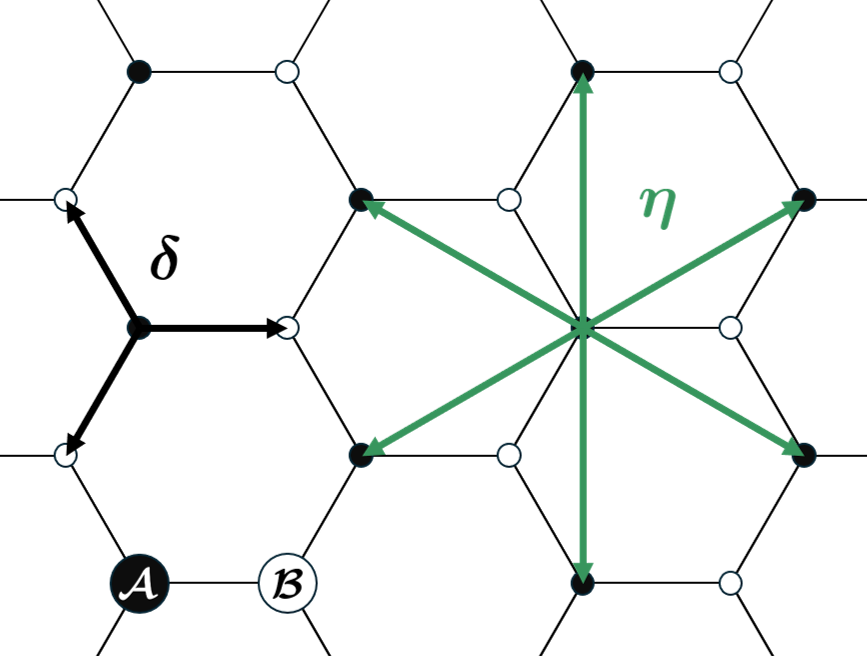}
    \caption{Position‐space vectors}
    \label{fig:position-vectors}
  \end{subfigure}
  \hfill
  \begin{subfigure}[b]{0.45\linewidth}
    \centering
    \includegraphics[width=.85\textwidth]{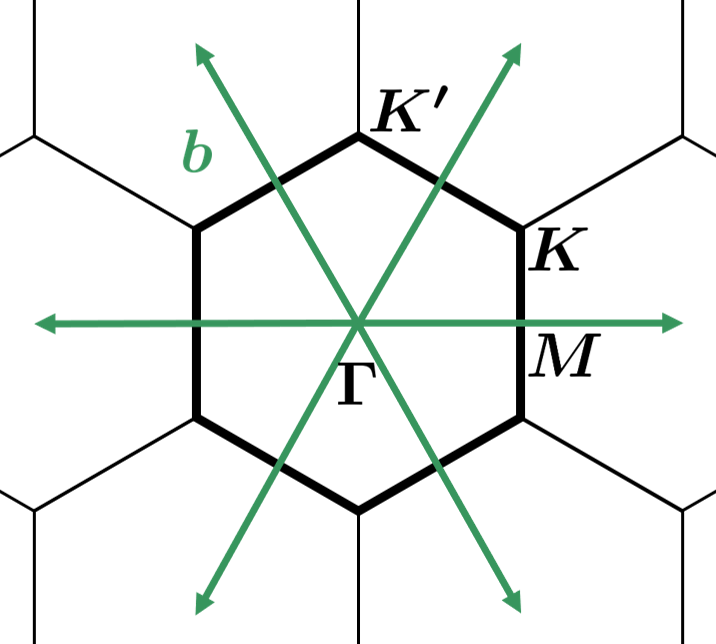}
    \caption{Momentum‐space vectors}
    \label{fig:momentum-vectors}
  \end{subfigure}
  \caption{Illustration of various position‐ and momentum‐space vectors defined in Table~\ref{tab:lattice}.}
  \label{fig:pos-mom-vectors}
\end{figure}

In this work, we consider a honeycomb lattice. Such a lattice is bipartite and we denote the first sublattice by $\mathcal{A}$ and the second one by $\mathcal{B}$. As illustrated in Fig.~\ref{fig:pos-mom-vectors}(a), every $\mathcal{A}$-site only neighbors $\mathcal{B}$-sites and vice versa. For a nearest-neighbor distance of $a$, the three  nearest-neighbor displacement vectors $\ddelta_i$ with $i\in\{1,2,3\}$ from sublattice $\mathcal{A}$ to sublattice $\mathcal{B}$ are $\big(a, 0\big)$, $\big(-\frac{a}{2}, \frac{\sqrt{3}}{2}a \big)$ and $\big(-\frac{a}{2}, -\frac{\sqrt{3}}{2}a \big)$. The lattice is chosen such that $\big(0, 0 \big) \in \mathcal{A}$. The reciprocal lattice with the first Brillouin zone as the central hexagon is shown in Fig.~\ref{fig:pos-mom-vectors}(b). The definitions of various (reciprocal) lattice vectors and characteristic lattice momenta are given in Table~\ref{tab:lattice}.

\begin{table}[H]
  \centering
  \begin{tabular}{llll}
    \textbf{Name} & \textbf{Definition} & \textbf{Exact form} & \textbf{Decimal with $a=1$} \\
    \hline
    $\ddelta_1$ & A $\to$ B NN & $(a,0)$ & $(1, 0)$ \\
    $\ddelta_2$ & A $\to$ B NN & $(-\tfrac{a}{2},\,\tfrac{\sqrt3a}{2})$ & $(-0.5, 0.866)$ \\
    $\ddelta_3$ & A $\to$ B NN & $(-\tfrac{a}{2},\,-\tfrac{\sqrt3a}{2})$ & $(-0.5,-0.866)$ \\ 
    \hline
    $\eeta_1$ & $\ddelta_1 - \ddelta_3$ & $(\tfrac{3a}{2},\,\tfrac{\sqrt3a}{2})$ & $(1.5,0.866)$ \\ 
    $\eeta_2$ & $\ddelta_2 - \ddelta_3$ & $(0,\,\sqrt3a)$ & $(0, 1.732)$ \\
    $\eeta_3$ & $\eeta_2 - \eeta_1$ & $(-\tfrac{3a}{2},\,\tfrac{\sqrt3a}{2})$ & $(-1.5,0.866)$ \\
    $\eeta_4$ & $-\eeta_1$ & $(-\tfrac{3a}{2},\,-\tfrac{\sqrt3a}{2})$ & $(-1.5,-0.866)$ \\ 
    $\eeta_5$ & $-\eeta_2$ & $(0,\,-\sqrt3a)$ & $(0, -1.732)$ \\
    $\eeta_6$ & $-\eeta_3$ & $(\tfrac{3a}{2},\,-\tfrac{\sqrt3a}{2})$ & $(1.5,-0.866)$ \\
    \hline \hline
    $\bb_1$ & $\eeta_i\cdot\bb_j=2\pi\delta_{ij}$ & $\frac{2\pi}{3a}(2,0)$ & $(4.189,0)$ \\
    $\bb_2$ & $\eeta_i\cdot\bb_j=2\pi\delta_{ij}$ & $\frac{2\pi}{3a}(1,\sqrt3)$ & $(2.094,3.628)$ \\
    $\bb_3$ & $\bb_2 - \bb_1$ & $\frac{2\pi}{3a}(-1,\sqrt3)$ & $(-2.094,3.628)$ \\
    $\bb_4$ & $-\bb_1$ & $\frac{2\pi}{3a}(-2,0)$ & $(-4.189,0)$ \\
    $\bb_5$ & $-\bb_2$ & $\frac{2\pi}{3a}(-1,-\sqrt3)$ & $(-2.094,-3.628)$ \\
    $\bb_6$ & $-\bb_3$ & $\frac{2\pi}{3a}(1,-\sqrt3)$ & $(2.094,-3.628)$ \\
    \hline
    $\GGamma$ & & $(0,0)$ & $(0,0)$ \\
    \hline
    $\KK_1$ & $(\bb_5+\bb_6)/3$ & $\frac{2\pi}{3\sqrt3a}(0,-2)$ & $(0,-2.418)$ \\
    $\KK_2$ & $(\bb_1+\bb_2)/3$ & $\frac{2\pi}{3\sqrt3a}(\sqrt3,1)$ & $(2.094,1.209)$ \\
    $\KK_3$ & $(\bb_3+\bb_4)/3$ & $\frac{2\pi}{3\sqrt3a}(-\sqrt3,1)$ & $(-2.094,1.209)$ \\
    \hline
    $\KK'_1$ & $-\KK_1$ & $\frac{2\pi}{3\sqrt3a}(0,2)$ & $(0,2.418)$ \\
    $\KK'_2$ & $-\KK_2$ & $\frac{2\pi}{3\sqrt3a}(-\sqrt3,-1)$ & $(-2.094,-1.209)$ \\
    $\KK'_3$ & $-\KK_3$ & $\frac{2\pi}{3\sqrt3a}(\sqrt3,-1)$ & $(2.094,-1.209)$ \\
    \hline
    $\MM_1$ & $\frac{1}{2}\bb_1$ & $\frac{\pi}{3a}(2,0)$ & $(2.094,0)$ \\
    $\MM_2$ & $\frac{1}{2}\bb_2$ & $\frac{\pi}{3a}(1,\sqrt3)$ & $(1.047, 1.814)$ \\
    $\MM_3$ & $\frac{1}{2}\bb_3$ & $\frac{\pi}{3a}(-1,\sqrt3)$ & $(-1.047, 1.814)$ \\
    $\MM_4$ & $\frac{1}{2}\bb_4$ & $\frac{\pi}{3a}(-2,0)$ & $(-2.094,0)$ \\
    $\MM_5$ & $\frac{1}{2}\bb_5$ & $\frac{\pi}{3a}(-1,-\sqrt3)$ & $(-1.047, -1.814)$ \\
    $\MM_6$ & $\frac{1}{2}\bb_6$ & $\frac{\pi}{3a}(1,-\sqrt3)$ & $(1.047, -1.814)$ \\
    \hline
  \end{tabular}
  \caption{Overview of various honeycomb lattice vectors and high-symmetry points of the Brillouin zone used throughout this work. NN stands for nearest-neighbor.}
  \label{tab:lattice}
\end{table}

\subsection{Tight-binding Hamiltonian in momentum space}
In order to describe the band structure of our system, mimicking a transition metal dichalcogenide (TMD), we employ the tight-binding Hamiltonian
\begin{equation}
    H_t = -t \sum_{\langle j, l \rangle}  \left( a^\dag_{j} b_{l} + b^\dag_{l} a_{j} \right) = -t \sum_{j} \sum_{\ddelta} \left( a^\dag_{j} b_{j+\ddelta} + b^\dag_{j+\ddelta} a_{j} \right),
\end{equation}
where $a_j^\dag$ and $b_j^\dag$ describe the creation of a fermion on lattice site $j$ on sublattice $\mathcal{A}$ and $\mathcal{B}$, respectively. Next we transform to momentum space using the definition \eqref{eq:fourierconvention}, to obtain
\begin{align}
    H_t &= -\frac{t}{N} \sum_j \sum_{\kk,\kk'} \sum_\ddelta \left( \e^{-\ii \kk \cdot \RR_j} a_\kk^\dag \e^{\ii \kk' \cdot (\RR_j+\ddelta)} b_{\kk'}  + \text{h.c.} \right) \\
    &= -\frac{t}{N} \sum_{\kk,\kk'} \sum_\ddelta \left[ a_\kk^\dag b_{\kk'} \e^{\ii \kk' \cdot \ddelta} \left( \sum_j \e^{\ii (\kk'-\kk) \cdot \RR_j} \right)  + \text{h.c.} \right] \\
    &= -t \sum_\kk \left[ a_\kk^\dag b_{\kk} \left( \sum_\ddelta \e^{\ii \kk \cdot \ddelta} \right) + \text{h.c.} \right] .
\end{align}
Defining $D_\kk = \sum_{\ddelta \in \mathcal{D}} \e^{\ii \kk \cdot \ddelta}$ with $\mathcal{D}=\left\{ \ddelta_1, \ddelta_2, \ddelta_3\right\}$, the Hamiltonian takes the form
\begin{equation}
    H_t = -t \sum_\kk \left( D_\kk a_\kk^\dag b_{\kk} + D^*_\kk b_\kk^\dag a_{\kk} \right).
\end{equation}
In matrix representation, this corresponds to
\begin{equation}
    H_t = \sum_\kk \left( a^\dag_\kk , b^\dag_\kk \right) \begin{pmatrix} 0 & -t D_\kk \\ -t D^*_\kk & 0\end{pmatrix} \begin{pmatrix} a_\kk \\ b_\kk \end{pmatrix}.
\end{equation}

\subsection{Band structure of the single-particle Hamiltonian}
Accounting for the energy offset $\Delta$ between $\mathcal{A}$ and $\mathcal{B}$ sites (see Eq.~(1) of the main text), the Hamiltonian becomes
\begin{equation}
    H_0 = \sum_\kk \left( a^\dag_\kk , b^\dag_\kk \right) \begin{pmatrix} \frac{\Delta}{2} & -t D_\kk \\ -t D^*_\kk & -\frac{\Delta}{2}\end{pmatrix} \begin{pmatrix} a_\kk \\ b_\kk \end{pmatrix}.
\end{equation}
Using that lattice momentum $\kk$ is conserved, we can diagonalize the $2\times2$ matrix to find the conduction and valence band eigenstates of the single-particle Hamiltonian. This leads to the dispersion relation
\begin{equation}
    E^{c/v}_\kk = \pm \sqrt{\left(\frac{\Delta}{2} \right)^2 + |tD_\kk|^2}
\end{equation}
of the two bands. The corresponding valence and conduction band creation operators are given by
\begin{align}
\begin{split}
v_{\kk}^\dag &= \mathcal{N}_\kk \left( a^\dag_\kk + \frac{\frac{\Delta}{2} + \sqrt{\left(\frac{\Delta}{2}\right)^2 + |t D_\kk|^2}}{tD_\kk} b^\dag_\kk  \right)  \eqqcolon P_\kk^* a^\dag_\kk + Q_\kk^* b^\dag_\kk \\
c_{\kk}^\dag &= \mathcal{N}_\kk \left(- \frac{\frac{\Delta}{2} + \sqrt{\left(\frac{\Delta}{2}\right)^2 + |t D_\kk|^2}}{tD_\kk^*} a^\dag_\kk + b^\dag_\kk  \right) \eqqcolon R_\kk^* a^\dag_\kk + S_\kk^* b^\dag_\kk,  
\end{split} \label{eq:vcinab}
\end{align}
where $\mathcal{N}_\kk$ ensures normalization. We have defined the complex parameters $P_\kk$, $Q_\kk$, $R_\kk$ and $S_\kk$ used below. With this, the Hamiltonian takes the form
\begin{equation}
    H_0 = \sum_\kk \left( E_\kk^c c^\dag_\kk c_\kk + E_\kk^v v^\dag_\kk v_\kk \right),
\end{equation}
given in the main text. As the transform is unitary, its inverse is simply given by
\begin{equation}
\begin{split}
a_{\kk}^\dag &= P_\kk v^\dag_\kk + R_\kk c^\dag_\kk \\
b_{\kk}^\dag &= Q_\kk v^\dag_\kk + S_\kk c^\dag_\kk.
\end{split} \label{eq:abinvc}    
\end{equation}

\section{Numerical solution of the ultracold atomic exciton problem} \label{sec:numericalsolution}
We begin with the single-particle Hamiltonian $H_0$ defined in Eq.~(1) in the main text, which for convenience we state here again
\begin{equation}
    H_0 = -t \sum_{\langle j, l \rangle}  \left( a^\dag_{j} b_{l} + b^\dag_{l} a_{j} \right) + \frac{\Delta}{2} \sum_{j} \left( a^\dag_{j} a_{j} - b^\dag_{j} b_{j} \right) \label{eq:HzerotightbindingSM}.
\end{equation}
The goal is to find the eigenstates $\ket{X_n}$ defined by $\aalpha_n$ corresponding to the variational exciton wave function Eq.~(3):
\begin{equation}
    \ket{X_n} = \sum_\pp \alpha_{n\pp} c^\dag_\pp v_{\pp}\ket{\text{FS}}_v \ket{0}_c . \label{eq:excitonstateSM}
\end{equation}
The basis most convenient for the following calculations is $a_\kk^\dag$, $b_\kk^\dag$ instead of valence and conduction band states. As a first step we thus transform Eqs. \eqref{eq:HzerotightbindingSM} and \eqref{eq:excitonstateSM} to this basis:
\begin{align}
    H_0 &= -t \sum_\kk \left(D_\kk a_\kk^\dag b_{\kk} + D^*_\kk b_\kk^\dag a_{\kk}\right) + \frac{\Delta}{2} \sum_\kk \left( a_\kk^\dag a_{\kk} - b_\kk^\dag b_{\kk} \right) \\
    \ket{X_n} &= \sum_\pp \alpha_{n\pp} \left(R_\pp^* a_\pp^\dag + S_\pp^* b_\pp^\dag \right) \left(P_\pp a_\pp + Q_\pp b_\pp \right) \prod_{\pp'} \left(P_{\pp'}^* a_{\pp'}^\dag + Q_{\pp'}^* b_{\pp'}^\dag \right) \ket{0}_a \ket{0}_b,
\end{align}
where the product over $\pp'$ runs over the full first Brillouin zone (as does the sum over $\pp$).
The index $n$ on $\alpha_{n\pp}$ will from now on be dropped for readability.

For the interaction, we begin with Eq.~(2) of the main text
\begin{equation}
    H_{\text{int}} = \sum_{jl} \bigg[\frac{1}{2} V^{(1)}_{jl} \left(a^\dag_{j}a^\dag_{l}a_{l}a_{j} + b^\dag_{j}b^\dag_{l}b_{l}b_{j}\right) + V^{(2)}_{jl} a^\dag_{j}b^\dag_{l}b_{l}a_{j} \bigg] \label{eq:HintrealspaceSM}.
\end{equation}
Assuming that the operators $a^\dag_j$ create particles in highly localized Wannier orbitals at lattice positions $\RR_j$, we can take $V_{jl}^{(1)} = \frac{d^2}{|\RR_l - \RR_j|^3}$ and $V_{jl}^{(2)} = \frac{d^2}{|\RR_l +\ddelta - \RR_j|^3}$ with $\ddelta$ one of the representatives translating from sublattice $\mathcal{A}$ to sublattice $\mathcal{B}$ (cf. Table~\ref{tab:lattice}). We can translate the sum over $l$ into a sum over relative positions between the two lattice sites $\rr \coloneqq \RR_l - \RR_j$. With this, $V^{(1)}$ and $V^{(2)}$ only depend on $\rr$, but not on $j$. Next, all operators are transformed to momentum space, leading to
\begin{equation}
    H_\text{int} = \frac{1}{N} \sum_{\kk \kk' \qq} \left[ \frac{1}{2} V_\qq^{(1)} \left(a_\kk^\dag a_{\kk'}^\dag a_{\kk'-\qq} a_{\kk+\qq} + b_\kk^\dag b_{\kk'}^\dag b_{\kk'-\qq} b_{\kk+\qq}\right) + V_\qq^{(2)} a_\kk^\dag b_{\kk'}^\dag b_{\kk'-\qq} a_{\kk+\qq} \right] \label{eq:Hintmomspace},
\end{equation}
with $V_\qq^{(1)} = \sum_\rr V_\rr^{(1)} \e^{-\ii \rr \cdot \qq}$ and $V_\qq^{(2)} = \sum_\rr V_\rr^{(2)} \e^{-\ii (\rr+\ddelta) \cdot \qq}$.

Having expressed $H=H_0+H_\text{int}$ in the $a_\kk^\dag$, $b_\kk^\dag$ basis, we calculate the energy of the exciton state $\bra{X_n} H \ket{X_n}$ subject to the condition $\braket{X_n|X_n} = 1$, as enforced by the energy functional 
\begin{equation}
    F[\aalpha] = \bra{X_n} H \ket{X_n} - \bra{\text{FS}} H \ket{\text{FS}} - E \braket{X_n|X_n}.
\end{equation}
From the stationary points given by $\frac{\delta F[\aalpha]}{\delta \alpha_\qq^*} = 0$, we obtain the eigenvalue equation (i.e. the Bethe-Salpeter equation), which is then solved numerically.

To obtain $F$, the following matrix elements are used
\begin{align}
    \bra{\text{FS}} H \ket{\text{FS}} &= \bra{\text{FS}} H_0 \ket{\text{FS}} + \bra{\text{FS}} H_\text{int} \ket{\text{FS}} \eqqcolon E_0 \\
    \bra{\text{FS}} H_0 \ket{\text{FS}} &= \sum_\kk E_\kk^v = - \sum_\kk \sqrt{\left(\frac{\Delta}{2}\right)^2 + |t D_\kk|^2} \\
    \bra{\text{FS}} H_\text{int} \ket{\text{FS}} &= \frac{1}{N} \sum_{\kk \kk'} \frac{1}{2} \left( |P_\kk|^2 |P_{\kk'}|^2 + |Q_\kk|^2 |Q_{\kk'}|^2 \right) \left( V_\zerovec^{(1)} - V_{\kk'-\kk}^{(1)} \right) + \frac{1}{N} \sum_{\kk \kk'} \left(|P_\kk|^2 |Q_{\kk'}|^2 V_\zerovec^{(2)} - P_\kk Q_{\kk'} P^*_{\kk'}Q_\kk^* V_{\kk'-\kk}^{(2)} \right) \nonumber \\
    &= \sum_{\kk \kk'} U_{\kk \kk'}^{(0)}
\end{align}
where $E_0$ denotes the (mean-field) ground state energy and $\ket{\text{FS}} = \ket{\text{FS}}_v \ket{0}_c$.

For the excited states, we have 
\begin{align}
    \bra{X_n} H \ket{X_n} &= \bra{X_n} H_0 \ket{X_n} + \bra{X_n} H_\text{int} \ket{X_n} \\
    \bra{X_n} H_0 \ket{X_n} &= \sum_\pp | \alpha_\pp |^2 E_\pp^c + \sum_\pp | \alpha_\pp |^2 \sum_{\kk \neq \pp} E_\kk^v \\
    \bra{X_n} H_\text{int} \ket{X_n} &= \sum_{\kk \kk'} \left( | \alpha_\kk|^2 U_{\kk \kk'}^{(1)} + \alpha^*_\kk \alpha_{\kk'} U_{\kk \kk'}^{(2)} + \alpha_{\kk'}^* \alpha_\kk U_{\kk \kk'}^{(3)} + |\alpha_{\kk'}|^2 U_{\kk \kk'}^{(4)} \right) + \sum_\pp |\alpha_\pp|^2 \sum_{\kk, \kk' \neq \pp} U_{\kk \kk'}^{(0)},
\end{align}
with
\begin{align}
    U_{\kk \kk'}^{(0)} &= \frac{1}{2N} \left( P_\kk P_{\kk'} P_{\kk'}^* P_\kk^* + Q_\kk Q_{\kk'} Q_{\kk'}^* Q_\kk^* \right) \left( V_\zerovec^{(1)} - V_{\kk'-\kk}^{(1)} \right) + 
    \frac{1}{N} \left(P_\kk Q_{\kk'}Q_{\kk'}^* P_\kk^* V_\zerovec^{(2)} - P_\kk Q_{\kk'} P^*_{\kk'}Q_\kk^* V_{\kk'-\kk}^{(2)} \right) \\
    U_{\kk \kk'}^{(1)} &= \frac{1}{2N} \left( R_\kk P_{\kk'} P_{\kk'}^* R_\kk^* + S_\kk Q_{\kk'} Q_{\kk'}^* S_\kk^* \right) \left( V_\zerovec^{(1)} - V_{\kk'-\kk}^{(1)} \right) + 
    \frac{1}{N} \left(R_\kk Q_{\kk'}Q_{\kk'}^* R_\kk^* V_\zerovec^{(2)} - R_\kk Q_{\kk'} P^*_{\kk'}S_\kk^* V_{\kk'-\kk}^{(2)} \right) \\
    U_{\kk \kk'}^{(2)} &= \frac{1}{2N} \left( R_\kk P_{\kk'} R_{\kk'}^* P_\kk^* + S_\kk Q_{\kk'} S_{\kk'}^* Q_\kk^* \right) \left( V_\zerovec^{(1)} - V_{\kk'-\kk}^{(1)} \right) + 
    \frac{1}{N} \left(R_\kk Q_{\kk'}S_{\kk'}^* P_\kk^* V_\zerovec^{(2)} - R_\kk Q_{\kk'} R^*_{\kk'}Q_\kk^* V_{\kk'-\kk}^{(2)} \right) \\
    U_{\kk \kk'}^{(3)} &= \frac{1}{2N} \left( P_\kk R_{\kk'} P_{\kk'}^* R_\kk^* + Q_\kk S_{\kk'} Q_{\kk'}^* S_\kk^* \right) \left( V_\zerovec^{(1)} - V_{\kk'-\kk}^{(1)} \right) + 
    \frac{1}{N} \left(P_\kk S_{\kk'}Q_{\kk'}^* R_\kk^* V_\zerovec^{(2)} - P_\kk S_{\kk'} P^*_{\kk'}S_\kk^* V_{\kk'-\kk}^{(2)} \right) \\
    U_{\kk \kk'}^{(4)} &= \frac{1}{2N} \left( P_\kk R_{\kk'} R_{\kk'}^* P_\kk^* + Q_\kk S_{\kk'} S_{\kk'}^* Q_\kk^* \right) \left( V_\zerovec^{(1)} - V_{\kk'-\kk}^{(1)} \right) + 
    \frac{1}{N} \left(P_\kk S_{\kk'}S_{\kk'}^* P_\kk^* V_\zerovec^{(2)} - P_\kk S_{\kk'} R^*_{\kk'}Q_\kk^* V_{\kk'-\kk}^{(2)} \right).
\end{align}
This yields
\begin{align}
    F[\aalpha] &= \bra{X_n} H \ket{X_n} - \bra{\text{FS}} H \ket{\text{FS}} - E \braket{X_n|X_n} \\
    &= \sum_\pp |\alpha_\pp|^2 \left( E_\pp^c - E_\pp^v \right) + \sum_{\kk \kk'} \left(| \alpha_\kk|^2 U_{\kk \kk'}^{(1)} + \alpha^*_\kk \alpha_{\kk'} U_{\kk \kk'}^{(2)} + \alpha_{\kk'}^* \alpha_{\kk} U_{\kk \kk'}^{(3)} + |\alpha_{\kk'}|^2 U_{\kk \kk'}^{(4)} \right) \nonumber \\&\qquad - \sum_\pp |\alpha_\pp|^2 \sum_{\kk \neq \pp} \left(U_{\kk \pp}^{(0)} + U_{\pp \kk}^{(0)} \right) - E \sum_\pp |\alpha_\pp|^2,
\end{align}
noting that $U_{\pp \pp}^{(0)}$ terms are zero, as seen in the definition of $U_{\kk \kk'}^{(0)}$. The stationary points are then given by
\begin{equation}
    0 = \frac{\delta F[\aalpha]}{\delta \alpha_\qq^*} = \alpha_\qq \left[ E_\qq^c - E_\qq^v + \sum_\kk \left( U_{\qq \kk}^{(1)} + U_{\kk \qq}^{(4)} - U_{\kk \qq}^{(0)} - U_{\qq \kk}^{(0)} \right) \right] + \sum_\kk \alpha_\kk \left( U_{\qq \kk}^{(2)} + U_{\kk \qq}^{(3)} \right) - \alpha_\qq E. \label{eq:finaleigvaleq}
\end{equation}
We implement \eqref{eq:finaleigvaleq} numerically using standard packages \cite{Numpy2020, Scipy2020, Matplotlib2007}, leading to the results for the eigenenergies and eigenstates given in the main text (see Figs.~2 and 3). To obtain stable results, the $\kk$ grid must both sample the entire Brillouin zone, in order to resolve states in the flat-band regime, and have very fine resolution near the K/K$^\prime$ points, in order to properly represent states in the effective mass regime. We therefore choose a two-part momentum grid. The first part of the grid is a regular triangular lattice, filling the hexagonal Brillouin zone; for the figures shown in the main text, we use 1200 uniform grid points. The second part is a logarithmic grid converging to the K/K$^\prime$ points from the six closest points of the triangular grid. The figures in the main text include 270 logarithmic steps to a minimum distance of 0.001 times the triangular-grid spacing. All lattice sums over this inhomogeneous grid are weighted using the local Voronoi cell areas.

\section{Effective mass approximation for weak interactions} \label{sec:effective}
In this section we present the derivation of the effective mass Schrödinger equation given as Eq.~(5) in the main text
\begin{equation}
    \left[- \frac{1}{m^*}\nabla_\rr^2 + V(\rr) \right] \psi(\rr) = E\psi(\rr)    \label{eq:effectivemassSM},
\end{equation}
which describes the system at weak interactions.
We begin with the full Hamiltonian of the system given by \eqref{eq:HzerotightbindingSM} and \eqref{eq:HintrealspaceSM}. In the limit of weak interactions, the interaction strength (e.g. quantified via the nearest-neighbor interaction strength $V_{NN}$, also called $V_1$ in the main text) is the smallest energy scale in the system, for example compared to $t$ and $\Delta$, but most importantly also when compared to the bandwidth $W = \sqrt{\left(\frac{\Delta}{2}\right)^2 + 9t^2} \ - \ \frac{\Delta}{2}$. The interaction \eqref{eq:HintrealspaceSM} is also the only off-diagonal contribution to our Hamiltonian, coupling states from valence and conduction band via intra- and interband transitions. In this limit the exciton will not extend over the full Brillouin zone, but only over a small region around the K/K$^\prime$ points. This fact naturally allows for further approximations that greatly simplify the Hamiltonian and lead to the effective mass approximation.

\subsection*{Simplifying the interaction}
As in the previous section, we begin by translating the interaction Hamiltonian to momentum space as given by \eqref{eq:Hintmomspace}
\begin{equation}
    H_\text{int} = \frac{1}{N} \sum_{\kk \kk' \qq} \left[ \frac{1}{2} V_\qq^{(1)} \left(a_\kk^\dag a_{\kk'}^\dag a_{\kk'-\qq} a_{\kk+\qq} + b_\kk^\dag b_{\kk'}^\dag b_{\kk'-\qq} b_{\kk+\qq}\right) + V_\qq^{(2)} a_\kk^\dag b_{\kk'}^\dag b_{\kk'-\qq} a_{\kk+\qq} \right].
\end{equation}
Using the unitary transform Eq.~\eqref{eq:abinvc}, we find
\begin{equation}
\begin{split}
    H_\text{int} = \frac{1}{N} \sum_{\kk \kk' \qq} \bigg[& v_\kk^\dag v_{\kk'}^\dag v_{\kk'-\qq} v_{\kk+\qq} \left(\frac{1}{2} V_\qq^{(1)} P_\kk P_{\kk'} P^*_{\kk'-\qq} P^*_{\kk+\qq} + \frac{1}{2} V_\qq^{(1)} Q_\kk Q_{\kk'} Q^*_{\kk'-\qq} Q^*_{\kk+\qq} + V_\qq^{(2)} P_\kk Q_{\kk'} Q^*_{\kk'-\qq} P^*_{\kk+\qq} \right) \\
    + & c_\kk^\dag v_{\kk'}^\dag c_{\kk'-\qq} v_{\kk+\qq} \left(\frac{1}{2} V_\qq^{(1)} R_\kk P_{\kk'} R^*_{\kk'-\qq} P^*_{\kk+\qq} + \frac{1}{2} V_\qq^{(1)} S_\kk Q_{\kk'} S^*_{\kk'-\qq} Q^*_{\kk+\qq} + V_\qq^{(2)} R_\kk Q_{\kk'} S^*_{\kk'-\qq} P^*_{\kk+\qq} \right) \\
    + & c_\kk^\dag v_{\kk'}^\dag v_{\kk'-\qq} c_{\kk+\qq} \left(\frac{1}{2} V_\qq^{(1)} R_\kk P_{\kk'} P^*_{\kk'-\qq} R^*_{\kk+\qq} + \frac{1}{2} V_\qq^{(1)} S_\kk Q_{\kk'} Q^*_{\kk'-\qq} S^*_{\kk+\qq} + V_\qq^{(2)} R_\kk Q_{\kk'} Q^*_{\kk'-\qq} R^*_{\kk+\qq} \right) \\
    + & v_\kk^\dag c_{\kk'}^\dag c_{\kk'-\qq} v_{\kk+\qq} \left(\frac{1}{2} V_\qq^{(1)} P_\kk R_{\kk'} R^*_{\kk'-\qq} P^*_{\kk+\qq} + \frac{1}{2} V_\qq^{(1)} Q_\kk S_{\kk'} S^*_{\kk'-\qq} Q^*_{\kk+\qq} + V_\qq^{(2)} P_\kk S_{\kk'} S^*_{\kk'-\qq} P^*_{\kk+\qq} \right) \\
    + & v_\kk^\dag c_{\kk'}^\dag v_{\kk'-\qq} c_{\kk+\qq} \left(\frac{1}{2} V_\qq^{(1)} P_\kk R_{\kk'} P^*_{\kk'-\qq} R^*_{\kk+\qq} + \frac{1}{2} V_\qq^{(1)} Q_\kk S_{\kk'} Q^*_{\kk'-\qq} S^*_{\kk+\qq} + V_\qq^{(2)} P_\kk S_{\kk'} Q^*_{\kk'-\qq} R^*_{\kk+\qq} \right) \\
    + & c_\kk^\dag c_{\kk'}^\dag c_{\kk'-\qq} c_{\kk+\qq} \left(\frac{1}{2} V_\qq^{(1)} R_\kk R_{\kk'} R^*_{\kk'-\qq} R^*_{\kk+\qq} + \frac{1}{2} V_\qq^{(1)} S_\kk S_{\kk'} S^*_{\kk'-\qq} S^*_{\kk+\qq} + V_\qq^{(2)} R_\kk S_{\kk'} S^*_{\kk'-\qq} R^*_{\kk+\qq} \right) \bigg] \label{eq:Hintmomspaceterms}. 
\end{split}
\end{equation}
Here, we have already used that the Hamiltonian has to conserve particle numbers in each band, as these band-wise particle numbers are fixed in both the Fermi sea $\ket{\text{FS}}$ at $(N,0)$ and the exciton states $\ket{X_n}$ at $(N-1,1)$. Using that, in the vicinity of the K/K$^\prime$ points, $D_\kk$ linearly approaches 0, one recognizes that $P_\kk$ and $S_\kk$ linearly approach 0 there as well. Expanding the prefactors of the various terms in Eq.~\eqref{eq:Hintmomspaceterms} in the distance to the K/K$^\prime$ points, we find only three terms with nonvanishing zeroth-order contributions (whose prefactors are approximately 1 in this region), leading to
\begin{equation}
    H_\text{int} = \frac{1}{N} \sum_{\kk \kk' \qq} \left[ \frac{1}{2} V_\qq^{(1)} \left(v_\kk^\dag v_{\kk'}^\dag v_{\kk'-\qq} v_{\kk+\qq} + c_\kk^\dag c_{\kk'}^\dag c_{\kk'-\qq} c_{\kk+\qq} \right) + V_\qq^{(2)} c_\kk^\dag v_{\kk'}^\dag v_{\kk'-\qq} c_{\kk+\qq} \right].
\end{equation}
Next we perform a particle-hole transform by defining $h_\qq^\dag \coloneqq v_{-\qq}$, which yields
\begin{equation}
\begin{split}
    H_\text{int} = \frac{1}{N} &\sum_{\kk \kk' \qq} \left[ \frac{1}{2} V_\qq^{(1)} \left(h_\kk^\dag h_{\kk'}^\dag h_{\kk'-\qq} h_{\kk+\qq} + c_\kk^\dag c_{\kk'}^\dag c_{\kk'-\qq} c_{\kk+\qq} \right) - V_\qq^{(2)} c_\kk^\dag h_{\kk'}^\dag h_{\kk'-\qq} c_{\kk+\qq} \right] \\
    & + \frac{1}{N} \sum_{\kk \kk'} \left[ \frac{1}{2} \left( V_\zerovec^{(1)} - V_{\kk-\kk'}^{(1)} \right) \left(1 - h_{\kk'}^\dag h_{\kk'} - h_{\kk}^\dag h_{\kk} \right)  + V_\zerovec^{(2)} c_{\kk}^\dag c_{\kk} \right].
\end{split}
\end{equation}
The first line contains the two-particle interaction terms that govern the dynamics of exciton formation, whereas the second line represents self-energy corrections. Below we show that the latter are momentum-independent within the Hilbert space of a single electron-hole pair.
To this end, we must show that $\sum_{\kk} V_{\kk-\kk'}^{(1)} = \text{const}$. From the definition of $V_\qq^{(1)}$ and $V_\qq^{(2)}$ (cf. Eq.~\eqref{eq:Hintmomspace}) we see that both are invariant under addition of a reciprocal lattice vector $\bb_j$. As the Brillouin zone is the set of all points closer to the origin than to any of the reciprocal lattice vectors, we can make the following argument:
Replacing $\kk$ by $\kk-\kk'$ shifts the Brillouin-zone-shaped summation boundary in momentum space. But as the interaction is periodic in reciprocal lattice vectors, all points that are now outside the original Brillouin zone precisely correspond to a point in the original Brillouin zone via one of the reciprocal lattice vectors. Thereby, identifying all terms of the sum of $\sum_\kk V_{\kk-\kk'}^{(1)}$ with a term of the sum $\sum_\kk V_{\kk}^{(1)}$, all sums are equal and independent of $\kk'$. Thus, representing just an energy shift, the second line can be dropped.

As a final step of simplifying the interaction, we use the knowledge of the structure of the Hilbert space given by \eqref{eq:excitonstateSM}, which contains exactly one conduction band particle and one valence band hole. This means the only nonzero contribution within this sector stems from the interaction between the bands, namely
\begin{equation}
    H_\text{int} = - \frac{1}{N} \sum_{\kk \kk' \qq} V_\qq^{(2)} c_\kk^\dag h_{\kk'}^\dag h_{\kk'-\qq} c_{\kk+\qq}. \\
\end{equation}
Here we already see the particle-hole transform aiding intuition: the effective interaction between particle and hole is attractive, even though all bare interactions are repulsive.

\subsection*{Effective mass approximation}
For a general derivation see Ref.~\cite{combescot2015excitons}.
The Hamiltonian consists of three terms: kinetic terms for the two bands and the interaction term we just derived:
\begin{equation}
    H = \sum_\kk \left( E_\kk^c c^\dag_\kk c_\kk - E_\kk^v h^\dag_\kk h_\kk \right) - \frac{1}{N} \sum_{\kk \kk' \qq} V_\qq^{(2)} c_\kk^\dag h_{\kk'}^\dag h_{\kk'-\qq} c_{\kk+\qq},
\end{equation}
where the constant shift appearing from the particle-hole transform of the kinetic terms has been dropped. In the vicinity of the band extrema at K/K$^\prime$, the band structure is approximately quadratic. Expanding in small momentum $\pp = \kk - \KK$ around $\KK$ (or $\KK'$) the dispersion becomes $\frac{\pp^2}{2m^*}$. Here $m^*$ represents the effective mass of a ``free" particle, which is determined by the curvature of the band structure at its minimum 
\begin{equation}
m^*_{c/v} = \left.\left(\frac{d^2 E^{c/v}_\kk}{d\kk^2}\right)^{-1}\right|_{\kk=\KK,\KK'} = \pm \frac{2 \Delta}{9 t^2 a^2},
\end{equation}
with $m^* \coloneqq m^*_c = - m^*_v$.
Furthermore, we can find a closed form for $V_\qq^{(2)}$ for small $\qq$ by performing a discrete Fourier transform of the real-space lattice and identifying the leading behavior. This yields
\begin{equation}
    V(\qq) = V_0 - \frac{2 \pi d^2}{A} |\qq| + \mathcal{O}(\qq^2),
\end{equation}
where $A=\frac{3 \sqrt{3}}{2}a^2$ is the real-space area of the Wigner-Seitz cell and $V_0 = \zeta_\triangle (3, \ddelta) \frac{d^2}{a^3} \approx 4.455 \frac{d^2}{a^3}$ with $\zeta_\triangle (3, \ddelta)$ the Epstein-zeta constant for a triangular lattice with offset $\ddelta$.

Since we know that only a small part of the Brillouin zone, i.e. $|\pp| \ll |\KK|$, contributes to the exciton state (this is also verified \textit{a posteriori} in the main text using the numerical exciton wave functions), we can restrict our momentum summation to this region, thereby yielding the Bethe-Salpeter equation (where $E$ is now measured from the band edge)
\begin{equation}
    \frac{\pp^2}{m^*} \alpha_\pp - \frac{1}{N}\sum_{\pp'} V(\pp'-\pp) \alpha_{\pp'} = E \alpha_\pp.
\end{equation}
Interestingly, all the remaining information about the lattice is now found in the interaction $V(\pp)$. Using a continuous Fourier transform to obtain $V(\rr)$ (absorbing the minus sign as in the main text), we can now translate the entire equation to real space, yielding the effective one-body Schrödinger equation for the relative motion
\begin{equation}
    \left[- \frac{1}{m^*}\nabla_\rr^2 + V(\rr) \right] \psi(\rr) = E\psi(\rr)
\end{equation}
with reduced mass $m^*/2$ in an attractive external potential $V(\rr)$ with $\psi(\rr)$ the continuous Fourier transform of $\alpha_\pp$. The potential $V(\rr)$ has the form $V(\rr) = -\frac{\dd^2}{|\rr|^3}$ for $|\rr|>r_0$ and $V(\rr) = -\frac{\dd^2}{r_0^3}$ for $|\rr|\le r_0$, with $r_0$ a short-distance cutoff. We solve this equation numerically by choosing an appropriate $\rr$-grid, expressing the Laplacian $\nabla_\rr^2$ via finite differences and subsequently diagonalizing the discretized Hamiltonian. This leads to the results presented in the main text.

\section{Effective flat-band Hamiltonian via Schrieffer-Wolff transformation} \label{sec:flat}
Here we will provide details on the derivation of the effective flat-band Hamiltonian Eq.~(6) given in the main text using a Schrieffer-Wolff transformation, leading to
\begin{equation}
    H_\text{FB} = \left(\Delta + \frac{6t^2}{\Delta} \right)+ \frac{2t^2}{\Delta} \sum_j \sum_\eeta d^\dag_{\rr_j} d_{\rr_j+\eeta} - V_{NN} \sum_{j} v^{(2)}_{\rr_j} d^\dag_{\rr_j} d_{\rr_j}. 
\end{equation}
We begin by revisiting the full Hamiltonian of the system, as given by \eqref{eq:HzerotightbindingSM} and \eqref{eq:HintrealspaceSM}. In the flat-band limit, the band gap $\Delta$ is the largest energy scale in the system. Therefore the energy offset term will be dominant in the Hamiltonian. As a consequence the hopping term will be effectively suppressed, as moving between $\mathcal{A}$ and $\mathcal{B}$ means bridging a very large energy gap. This causes the Hilbert space to effectively decouple into two subspaces associated with the two sublattices. On a formal level, we will decouple these two subspaces kinetically using a Schrieffer-Wolff transformation and derive an effective model for low-energy states in this parameter regime.

\subsection*{Schrieffer-Wolff transformation}
First we may separate the full Hamiltonian $H$ into the diagonal part (dividing by $\Delta$ to obtain a dimensionless form)
\begin{equation}
    H_d = \frac{1}{2} \sum_{j} \left( a^\dag_{j} a_{j} - b^\dag_{j} b_{j} \right) + \frac{1}{\Delta} \sum_{jl} \bigg[\frac{1}{2} V^{(1)}_{jl} \left(a^\dag_{j}a^\dag_{l}a_{l}a_{j} + b^\dag_{j}b^\dag_{l}b_{l}b_{j}\right) + V^{(2)}_{jl} a^\dag_{j}b^\dag_{l}b_{l}a_{j} \bigg]
\end{equation}
and a perturbation
\begin{equation}
    Y = -\frac{t}{\Delta} \sum_{\langle j, l \rangle}  \left( a^\dag_{j} b_{l} +  b^\dag_{l} a_{j} \right),
\end{equation}
with $H = H_d + Y$.
The goal is now to rotate the full Hamiltonian $H$ using a unitary transformation generated by an operator $S$, such that the two energetically different subspaces are decoupled using $\tilde{H} = \e^S H \e^{-S}$. Assuming $S$ to be small, we can expand $\tilde{H}$ using the Baker-Campbell-Hausdorff formula to arrive at
\begin{align}
    \tilde{H} &= H + \comm{S}{H} + \frac{1}{2} \comm{S}{\comm{S}{H}} + \frac{1}{6}\comm{S}{\comm{S}{\comm{S}{H}}} + \cdots \label{eq:BCHshort} \\
    \tilde{H} &= H_d + Y + \comm{S}{H_d} + \comm{S}{Y} + \frac{1}{2} \comm{S}{\comm{S}{H_d}} + \frac{1}{2} \comm{S}{\comm{S}{Y}} + \frac{1}{6}\comm{S}{\comm{S}{\comm{S}{H_d}}} + \frac{1}{6}\comm{S}{\comm{S}{\comm{S}{Y}}} + \cdots \label{eq:BCHlong}
\end{align}
For the perturbative expansion to be carried out, we define dimensionless parameters $\epsilon_t=\frac{t}{\Delta}$ and $\epsilon_v=\frac{V_{NN}}{\Delta}$ where $V_{NN}$ is the nearest-neighbor interaction strength of the dipolar potential and we introduce $v_{ij}^{(l)} = \frac{V_{ij}^{(l)}}{V_{NN}}$. We assume $\epsilon_t \ll 1$ and $\epsilon_v \ll 1$ without any explicit assumption about their relative magnitude.
The Hamiltonian and the generator $S$ are split into parts as defined by their scaling with the small parameters $\epsilon_t$ and $\epsilon_v$,
\begin{align}
    H_d &= H_g + \epsilon_v H_\text{int} \\
    &\quad H_g = \frac{1}{2} \sum_{j} \left(a^\dag_{j} a_{j} - b^\dag_{j} b_{j}\right) \notag \\
    &\quad H_\text{int} = \sum_{jl} \left[ \frac{1}{2} v^{(1)}_{jl} \left(a^\dag_{j}a^\dag_{l}a_{l}a_{j} + b^\dag_{j}b^\dag_{l}b_{l}b_{j}\right) + v^{(2)}_{jl} a^\dag_{j}b^\dag_{l}b_{l}a_{j} \right] \notag\\
    Y &=\epsilon_t H_t \\
    &\quad H_t = - \sum_{\langle j l \rangle} \left( a^\dag_{j} b_{l} +  b^\dag_{l} a_{j} \right) \notag \\
    S &= \epsilon_t S_1 + \epsilon_t \epsilon_v S_2 \\
    &\quad S_1 = - \sum_{\langle j l \rangle} \left( a^\dag_{j} b_{l} -  b^\dag_{l} a_{j} \right) \notag \\
    &\quad S_2 = \sum_{\langle j l \rangle} \sum_m \left[ \left( v_{ml}^{(2)} - v_{mj}^{(1)} \right) a_m^\dag \left( b_l^\dag a_j - a_j^\dag b_l \right) a_m- \left( v_{jm}^{(2)} - v_{lm}^{(1)} \right) b_m^\dag \left( b_l^\dag a_j - a_j^\dag b_l \right) b_m \right] \notag.
\end{align}
As can be seen, $S$ comprises a hopping term $S_1$ and an interaction-induced hopping term $S_2$. In order to decouple the two subspaces in $\tilde{H}$, the off-diagonal contribution $Y$ has to be canceled. This is done by the first commutator $\comm{S}{H_d}$ in Eq.~\eqref{eq:BCHlong}, which evaluates to 
\begin{equation}
    \comm{S}{H_d} = - \epsilon_t H_t + \epsilon_t \epsilon_v^2 Z, \label{eq:schriefferwolffcondition}
\end{equation}
where $Z$ is an $\mathcal{O}(1)$ term with respect to $\epsilon_t$ and $\epsilon_v$. In practice, $S_1$ was found as the solution to $\comm{S_1}{H_g} = - H_t$ and $S_2$ was chosen in order to cancel the off-diagonal terms of order $\epsilon_t \epsilon_v$, ensuring $\comm{S_2}{H_g} = -\comm{S_1}{H_\text{int}}$ and leaving $Z=\comm{S_2}{H_\text{int}}$.
We can expand the Baker-Campbell-Hausdorff expression \eqref{eq:BCHshort} to third order in the small parameters $\epsilon_t$ and $\epsilon_v$. After cancellations following from \eqref{eq:schriefferwolffcondition} have been made one finds
\begin{equation}
    \tilde{H} = H_g + \epsilon_v H_\text{int} + \frac{1}{2} \epsilon_t^2 \comm{S_1}{H_t} + \epsilon_t \epsilon_v^2 Z + \frac{1}{2} \epsilon_t^2 \epsilon_v \comm{S_2}{H_t} + \frac{1}{3} \epsilon_t^3 \comm{S_1}{\comm{S_1}{H_t}} + \mathcal{O}(\epsilon^4).
\end{equation}
Here $\comm{S_1}{H_t} = 2\sum_{\langle\langle j, l \rangle\rangle} \left(a^\dag_{j} a_{l} -  b^\dag_{j} b_{l}\right)$, where $\langle\langle \cdot , \cdot \rangle\rangle$ denotes pairs of sites on the same sublattice connected via two nearest-neighbor hops, explicitly including the return paths. Ordering  the contributions by their relative size, the leading terms are: (i) the sublattice energy offset $H_g$, (ii) the particle-particle interaction $H_\text{int}$, and (iii) the double-hopping processes $\comm{S_1}{H_t}$ that preserve sublattice occupation.  
At third order, we encounter $Z$ and $\comm{S_2}{H_t}$, which describe interaction-assisted hopping processes involving six operators, e.g. $\sum_{ij\langle kl\rangle} v^{(2)}_{jk} v^{(1)}_{il}\, a^\dag_k b^\dag_i b^\dag_j b_j b_i b_l$ or $\sum_{ijk\langle lm\rangle} v^{(2)}_{ij} \, a^\dag_i b^\dag_j b^\dag_l a_m b_j b_k$ , respectively, as well as $\comm{S_1}{\comm{S_1}{H_t}}$, which represents triple-hopping. These processes are already parametrically suppressed compared to the leading three contributions. Furthermore, $Z$ and $\comm{S_1}{\comm{S_1}{H_t}}$ both transfer particles between sublattices -- the very transitions that the unitary rotation was designed to eliminate, since such processes are strongly suppressed by the large sublattice energy splitting $H_g$. 

For these reasons, all third-order terms can be consistently neglected if $\epsilon_v, \epsilon_t \ll 1$. The small deviation of the $X_{\GGamma}$ energy between the full calculation and the flat-band model for intermediate values of $\frac{\Delta}{t} = \epsilon_t^{-1}$ in Fig.~2 of the main text can be attributed to the absence of scale separation from the term $\frac{1}{2} \epsilon_t^2 \epsilon_v \comm{S_2}{H_t}$. This disparity vanishes as expected with decreasing $\epsilon_t$.

The remaining Hamiltonian $\tilde{H}$ then reads
\begin{equation}
    \tilde{H} = \frac{1}{2} \sum_{j}\left( a^\dag_{j} a_{j} - b^\dag_{j} b_{j}\right) + \frac{V_{NN}}{\Delta} \sum_{jl}\left[ \frac{1}{2} v^{(1)}_{jl} \left(a^\dag_{j}a^\dag_{l}a_{l}a_{j} + b^\dag_{j}b^\dag_{l}b_{l}b_{j}\right) + v^{(2)}_{jl} a^\dag_{j}b^\dag_{l}b_{l}a_{j}\right] + \frac{t^2}{\Delta^2} \sum_{\langle\langle j, l \rangle\rangle}\left( a^\dag_{j} a_{l} -  b^\dag_{j} b_{l}\right).
\end{equation}

Next, we will use knowledge about the relevant Hilbert space of the exciton problem: both $\ket{\text{FS}}$ and $\ket{X_n}$ are defined via conduction and valence band operators $c^\dag_\kk$ and $v^\dag_\kk$. From the transform \eqref{eq:vcinab} to $a^\dag_\kk$ and $b^\dag_\kk$ we find that for $\Delta\gg t$ (or $\epsilon_t \ll 1$) the two bands coincide with the two sublattices, i.e. $c^\dag_\kk \approx a^\dag_\kk$ and $v^\dag_\kk \approx b^\dag_\kk$. As the valence band, which approaches the $\mathcal{B}$ sublattice in this limit, will be (almost) completely filled, we perform the particle-hole transform $h^\dag_i \coloneqq b_i$ to arrive at
\begin{equation}
    \tilde{H} = \frac{1}{2} \sum_{j}\left( a^\dag_{j} a_{j} + h^\dag_{j} h_{j}\right) + \frac{V_{NN}}{\Delta} \sum_{jl}\left[ \frac{1}{2} v^{(1)}_{jl} \left(a^\dag_{j}a^\dag_{l}a_{l}a_{j} + h^\dag_{j}h^\dag_{l}h_{l}h_{j}\right) - v^{(2)}_{jl} a^\dag_{j}h^\dag_{l}h_{l}a_{j}\right] + \frac{t^2}{\Delta^2} \sum_{\langle\langle j, l \rangle\rangle}\left( a^\dag_{j} a_{l} +  h^\dag_{j} h_{l}\right) + C,
\end{equation}
where $C = \frac{V_{NN}}{\Delta}\sum_{jl} \left(\frac{1}{2} v^{(1)}_{jl} - v^{(1)}_{jl} h^\dag_{j}h_{j} + v^{(2)}_{jl} a^\dag_{j}a_{j} \right)$ only depends on the total occupation number of the sublattices due to translational symmetry. As we know from the exciton ansatz \eqref{eq:excitonstateSM}, we have exactly one hole in the valence band and one particle in the conduction band. Therefore we can drop the $v^{(1)}_{jl}$-dependent terms of the translated Hamiltonian $\tilde{H}$. The constant energy shift $C$ does not impact the physics and will therefore be dropped as well. 
Lastly, we reintroduce the factor of $\Delta$ we removed earlier and arrive at the flat-band Hamiltonian
\begin{equation}
    H_\text{FB} = \frac{\Delta}{2} \sum_{j}\left( a^\dag_{j} a_{j} + h^\dag_{j} h_{j}\right) - V_{NN} \sum_{jl} v^{(2)}_{jl} a^\dag_{j}h^\dag_{l}h_{l}a_{j} + \frac{t^2}{\Delta} \sum_{\langle\langle j, l \rangle\rangle}\left( a^\dag_{j} a_{l} +  h^\dag_{j} h_{l}\right). \label{eq:flatbandhamiltonianSM}
\end{equation}

\subsection*{Effective single-particle Hamiltonian}
From \eqref{eq:flatbandhamiltonianSM} and the exciton ansatz \eqref{eq:excitonstateSM} we see that, while we describe a two-particle system, the only relevant degree of freedom is the relative position between the electron and the hole. We shall now use the zero-momentum condition of our ansatz to translate the flat-band Hamiltonian into a single-particle problem. To this end, we first separate $H_\text{FB}$ into the quadratic part $H_\text{FB,1}$ and the quartic part $H_\text{FB,2}$. We rewrite the two-step hopping sum $\sum_{\langle\langle j, l \rangle\rangle}$ using the six sublattice-neighbors $\eeta$ (cf. Fig.~\ref{fig:pos-mom-vectors}(a) and Table~\ref{tab:lattice}) with $\sum_\ddelta \sum_{\ddelta'} f(\ddelta-\ddelta') = 3f(0) + \sum_\eeta f(\eeta)$.

\vspace{0.15cm}
We thus arrive at
\begin{align}
    H_\text{FB,1} &= \frac{\Delta}{2} \left(1 + \frac{6t^2}{\Delta^2}\right) \sum_{j}\left( a^\dag_{j} a_{j} + h^\dag_{j} h_{j}\right)  + \frac{t^2}{\Delta} \sum_j \sum_\eeta \left( a^\dag_{j} a_{j+\eeta} + h^\dag_{j} h_{j+\eeta}\right) \\
    &= \frac{\Delta}{2} \left(1 + \frac{6t^2}{\Delta^2}\right) \frac{1}{N} \sum_{j} \sum_{\kk \kk'} \e^{\ii \RR_j\cdot(\kk-\kk')} \left( a^\dag_{\kk'} a_\kk + h^\dag_{\kk'} h_\kk \right) + \frac{t^2}{\Delta} \frac{1}{N} \sum_j \sum_\eeta \sum_{\kk \kk'} \e^{\ii \RR_j\cdot(\kk-\kk')} \e^{\ii \eeta \cdot \kk} \left( a^\dag_{\kk'} a_\kk + h^\dag_{\kk'} h_\kk \right).
\end{align}
We evaluate the Kronecker deltas $\delta_{\kk \kk'}$ arising from the summation over the exponentials and, defining constant energy shift $E_g = \frac{\Delta}{2} \left(1 + \frac{6t^2}{\Delta^2}\right)$ and band structure modification $C_\kk = \frac{t^2}{\Delta} \sum_\eeta \e^{\ii \eeta \cdot \kk}$, we obtain the simplified Hamiltonian
\begin{equation}
    H_\text{FB,1} = \sum_\kk \left( E_g + C_\kk \right) \left( a^\dag_{\kk} a_\kk + h^\dag_{\kk} h_\kk \right).
\end{equation}

We now consider again the Hilbert space spanned by exciton ansatz \eqref{eq:excitonstateSM}, consisting of zero-total-momentum states in which an electron and a hole are created with opposite momenta. It is thus convenient to define a pair creation operator $A^\dag_\pp \coloneqq c^\dag_\pp h^\dag_{-\pp}$ and rewrite the exciton ansatz as
\begin{align}
    \ket{X_n} &= \sum_\pp \alpha_{n\pp} c^\dag_\pp h^\dag_{-\pp} \ket{0}_c \ket{0}_h \\
    &= \sum_\pp \alpha_{n\pp} A^\dag_\pp \ket{0}_A.
\end{align}
With respect to the Hilbert space spanned by $A^\dag_\pp \ket{0}_A$, an equivalent representation of $H_\text{FB,1}$ (within the single-pair subspace) is given by
\begin{align}
    H_\text{FB,1} &= \sum_\kk 2 \left( E_g + C_\kk \right) A^\dag_\kk A_\kk \\
    &= 2E_g + \frac{2t^2}{\Delta} \sum_\kk \left( \sum_\eeta \e^{\ii \eeta \cdot \kk} \right) A^\dag_\kk A_\kk \\
    &= \left(\Delta + \frac{6t^2}{\Delta} \right)+ \frac{2t^2}{\Delta} \sum_j \sum_\eeta d^\dag_{\rr_j} d_{\rr_j+\eeta}.
\end{align}
Here $d^\dag_{\rr_j} = \frac{1}{\sqrt{N}} \sum_\kk \e^{-\ii \kk \cdot \rr_j} A_\kk^\dag$ denotes the creation operator in the relative-position lattice defined by the vectors connecting the electron to the hole. As we chose $\big(0, 0 \big) \in \mathcal{A}$, this relative-position lattice is identical to the sublattice $\mathcal{B}$. 

We now turn to expressing $H_\text{FB,2}$ and $\ket{X_n}$ in terms of $d_{\rr_j}^\dag$ as well. In the case of $H_{\text{FB},2}$, we make use of the fact that $v^{(2)}_{jl}$ depends only on the distance $|\RR_l + \ddelta - \RR_j|$, and accordingly introduce the relabeling $\RR \coloneqq \RR_j$ and $\rr \coloneqq \RR_l - \RR_j$. We then have
\begin{align}
    H_\text{FB,2} &= - V_{NN} \sum_{jl} v^{(2)}_{jl} a^\dag_{j}h^\dag_{l}h_{l}a_{j} \\
    &= - V_{NN} \sum_{\RR \rr} v^{(2)}_{\rr} a^\dag_{\RR}h^\dag_{\RR+\rr+\ddelta}h_{\RR+\rr+\ddelta}a_{\RR} \\
    &= - \frac{V_{NN}}{N} \sum_{\kk \kk' \qq} v^{(2)}_\qq a^\dag_{\kk}h^\dag_{\kk'}h_{\kk'-\qq}a_{\kk+\qq},
\end{align}
where we defined $v^{(2)}_\qq \coloneqq \sum_\rr v^{(2)}_{\rr} \e^{-\ii (\rr+\ddelta) \cdot \qq}$. As before, we can restrict the Hamiltonian to the parts that remain within the zero-total-momentum Hilbert space, which allows to eliminate the sum over $\kk'$. Hence
\begin{align}
    H_\text{FB,2} &= - \frac{V_{NN}}{N} \sum_{\kk \qq} v^{(2)}_\qq a^\dag_{\kk}h^\dag_{-\kk}h_{-\kk-\qq}a_{\kk+\qq} \\
    &= - \frac{V_{NN}}{N} \sum_{\kk \qq} v^{(2)}_\qq A^\dag_\kk A_{\kk+\qq} \\
    &= - \frac{V_{NN}}{N^2} \sum_{\kk \qq} v^{(2)}_\qq \sum_{jl} \e^{-\ii \rr_j \cdot \kk} \e^{\ii \rr_l \cdot (\kk+\qq)} d^\dag_{\rr_j} d_{\rr_l}\\
    &= - \frac{V_{NN}}{N}  \sum_{j} d^\dag_{\rr_j} d_{\rr_j}  \left(\sum_{\qq} v^{(2)}_\qq \e^{\ii \rr_j \cdot \qq} \right).
\end{align}
Using $\sum_{\qq} v^{(2)}_\qq \e^{\ii \rr_j \cdot \qq} = \sum_{\qq} \sum_\rr v^{(2)}_{\rr} \e^{-\ii (\rr+\ddelta) \cdot \qq} \e^{\ii \rr_j \cdot \qq} = N v^{(2)}_{\rr_j}$, we arrive at the final single-particle Hamiltonian
\begin{equation}
    H_\text{FB} = \left(\Delta + \frac{6t^2}{\Delta} \right)+ \frac{2t^2}{\Delta} \sum_j \sum_\eeta d^\dag_{\rr_j} d_{\rr_j+\eeta} - V_{NN} \sum_{j} v^{(2)}_{\rr_j} d^\dag_{\rr_j} d_{\rr_j}, \label{eq:flatbandhamiltoniansingleparticleSM}
\end{equation}
where all lattice vectors $\rr_j$ now live in the relative-position lattice, which is identical to the $\mathcal{B}$ sublattice.
The corresponding exciton ansatz is given by
\begin{equation}
    \ket{X_n} = \sum_j \beta_{n,\rr_j} d^\dag_{\rr_j} \ket{0}_d \label{eq:flatbandsingleparticlestateSM}
\end{equation}
with $\beta_{n,\rr_j} = \frac{1}{\sqrt{N}} \sum_\pp \alpha_{n\pp} \e^{\ii \pp \cdot \rr_j}$. The effective Schrödinger equation $H_\text{FB}\ket{X_n}=E\ket{X_n}$ is then solved numerically to produce Fig.~2 and Fig.~3 in the main text.

\subsection*{Three-level system}
Apart from the constant energy shift, the Hamiltonian \eqref{eq:flatbandhamiltoniansingleparticleSM} has two terms, the second-order kinetic hopping term scaling with $\frac{2t^2}{\Delta}$ and the interaction term scaling with $V_{NN}$. In the limit of large $\Delta$, or more precisely small $\frac{2t^2}{\Delta V_{NN}}$, the interaction term separates the Hilbert space into energetically distinct sectors that cannot be mixed by hopping to a relevant extent. 

\begin{figure}
    \centering
    \includegraphics[width=0.4\linewidth]{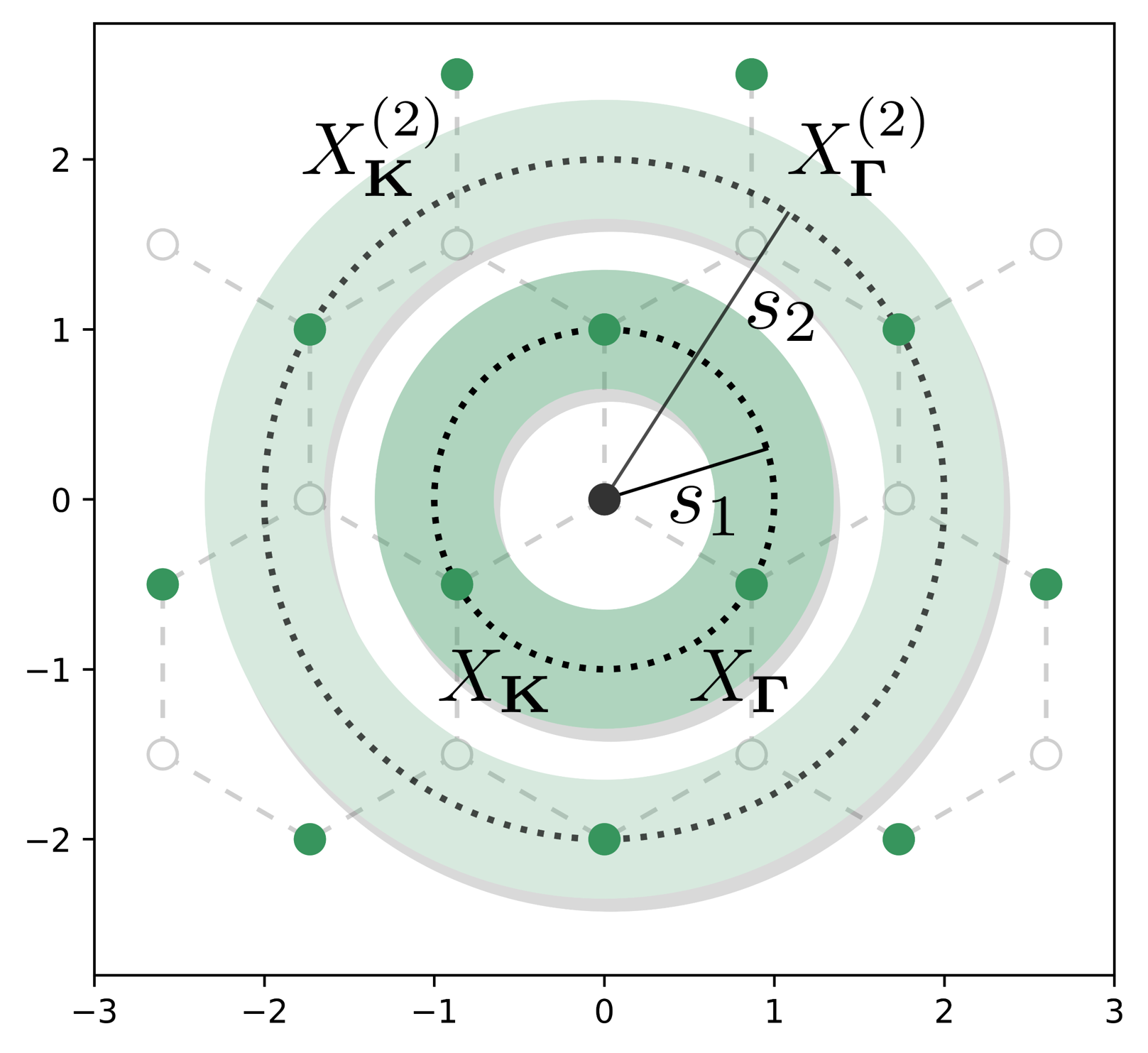}
    \caption{\justifying \textbf{Illustration of the emerging decoupled three-level systems in the flat-band limit.} Reproduction of Fig.~3(d) from the main text with added axes and distances. The relative-position lattice is shown in green.}
    \label{fig:threelevelSM}
\end{figure}

Denoting the shortest and second-shortest vector lengths by $s_1$ and $s_2$, respectively, we can rewrite the low-energy Hamiltonian as
\begin{equation}
    H_\text{FB} = \left(\Delta + \frac{6t^2}{\Delta} \right)+ \frac{2t^2}{\Delta} \sum_j \sum_\eeta d^\dag_{\rr_j} d_{\rr_j+\eeta} - V_{NN} \left[ \sum_{|\rr_j|=s_1} v^{(2)}_{s_1} d^\dag_{\rr_j} d_{\rr_j} + \sum_{|\rr_j|=s_2} v^{(2)}_{s_2} d^\dag_{\rr_j} d_{\rr_j} + \cdots \right].
\end{equation}
For our triangular relative-coordinate lattice with an offset (see Fig.~\ref{fig:threelevelSM}), we have $s_1=1$, $s_2=2$ in units of the nearest-neighbor distance $a$, leading to $v^{(2)}_{s_1}=1$ and $v^{(2)}_{s_2}=\frac{1}{8}$ for dipolar interaction. For $\frac{2t^2}{\Delta}\ll\left(v^{(2)}_{s_1}-v^{(2)}_{s_2} \right) V_{NN} =\frac78 V_{NN}$, the three shortest lattice vectors define the relevant basis states, leading to an emergent three-level system given in the main text 
\begin{equation}
    H_3 = \left(\Delta + \frac{6t^2}{\Delta} \right)+ \frac{2t^2}{\Delta} \sum_{\substack{j \neq l\\ |\rr_j|,|\rr_l|=1}} d^\dag_{\rr_j} d_{\rr_l} - V_{NN} \sum_{|\rr_j|=1}  d^\dag_{\rr_j} d_{\rr_j},
\end{equation}
or, given in matrix form as in the main text (up to constant contributions),
\begin{equation}
    H_3 = \left(\Delta + \frac{6t^2}{\Delta} - V_{NN}\right) \mathbf{1}_3 + \frac{2t^2}{\Delta} \begin{pmatrix} 0 & 1 & 1 \\ 1 & 0 & 1 \\ 1 & 1 & 0 \end{pmatrix}.
\end{equation}
This simple three-level system (with its corresponding eigenstates called $X_\KK$ and $X_\GGamma$ in the main text) with the matrix spectrum $\frac{2t^2}{\Delta}\{-1,-1,2\}$ asymptotically reproduces both exciton energies and eigenstates in the flat-band limit, specifically when $\frac{2t^2}{\Delta V_{NN}}$ is small.

The next three-level system (with its eigenstates called $X^{(2)}_\KK$ and $X^{(2)}_\GGamma$ in the main text) can be found by repeating the argument for the $s_2$ ring.

\section{Fermi's golden rule in band basis} \label{sec:fermi}
In this section we present details on the derivation of the perturbation operator $\hat{O}$ appearing in Eq.~(7) in the main text, which enters Fermi's golden rule 
\begin{equation}
    A(\omega) = 2 \pi \sum_n \left| \bra{X_n}\hat{O} \ket{\text{FS}} \right|^2 \delta(E_n - E_0 - \omega). \label{eq:fermisgoldenruleSM}
\end{equation}

\begin{figure}[h]
    \centering
    \includegraphics[width=0.7\linewidth]{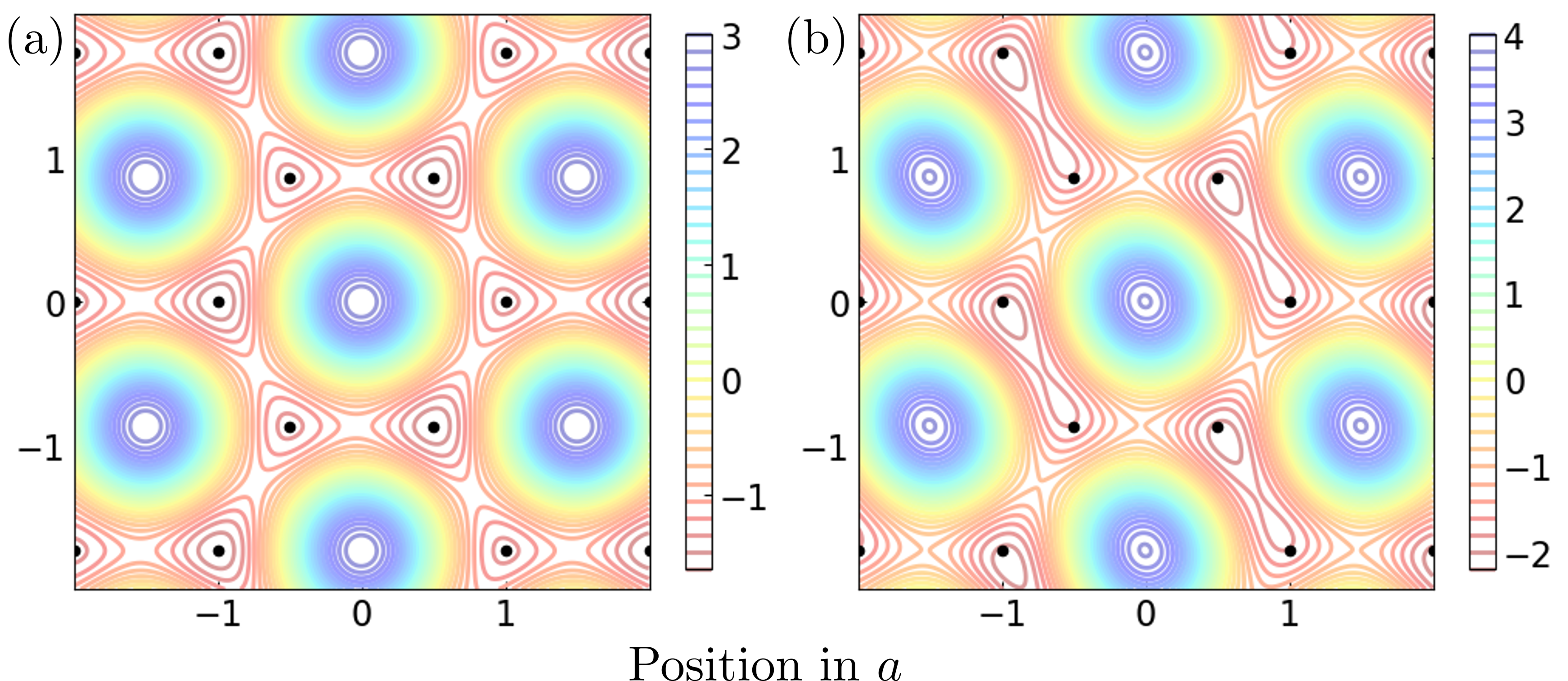}
    \caption{\justifying \textbf{External honeycomb lattice potential.} Potential generated by three interfering laser beams intersecting with an angle of $120^\circ$. The offset between $\mathcal{A}$-sites and $\mathcal{B}$-sites (both shown in black) is caused by the polarization of the light field here, but can also be realized using a triangular superlattice potential. (a) An equal intensity $I_1=I_2=I_3$ of all incident laser beams leads to a $C_3$-symmetric potential field, (b) unequal laser intensities $I_1=I_2=\frac{I_3}{2}$ break the symmetry, manifesting as asymmetric hopping in the three nearest-neighbor directions $\mathcal{D}$.}
    \label{fig:externalpotentialsSM}
\end{figure}

We begin by describing the external potential: an optical honeycomb lattice is created by three laser beams interfering at an angle of $120^\circ$ with in-plane polarization \cite{Flaeschner2018}. For laser intensities $I_1$, $I_2$ and $I_3$, the laser potential is given by
\begin{equation}
    V_\text{hex}(\rr) \propto V_{12} \cos \left( \rr \cdot \bb_3 \right) + V_{23} \cos \left( \rr \cdot \bb_1 \right)+V_{13} \cos \left( \rr \cdot \bb_2 \right),
\end{equation}
with $V_{ij} = \sqrt{I_iI_j}$. Additionally, neighboring sites of the lattice experience alternating local polarizations of the light field, which couple to the spins \cite{soltan2011multi}, leading to a sublattice-dependent potential term $V_\text{off}$ of the form
\begin{equation}
    V_\text{off} \propto V_{12} \sin \left( \rr \cdot \bb_3 \right) + V_{23} \sin \left( \rr \cdot \bb_1 \right)+V_{13} \sin \left( \rr \cdot \bb_2 \right).
\end{equation}
The polarization potential is equal for all fermions, as the spins are already aligned perpendicular to the lattice plane. The offset for ultracold molecules can be created using a superlattice potential. The total external potential $V_\text{ext}=V_\text{hex}+V_\text{off}$ is shown in Fig.~\ref{fig:externalpotentialsSM}(a).

Next, we introduce sublattice-resolved Bloch basis states $\ket{\psi_{\qq, a}}$ ($\ket{\psi_{\qq, b}}$) for the sublattices $\mathcal{A}$ ($\mathcal{B}$); these are the states created by $a_\qq^\dag$ ($b_\qq^\dag$). Via Fourier transform, we arrive at a local basis of Wannier states
\begin{equation}
\begin{split}
    \ket{w_{j,a}} &= \frac{1}{\sqrt{N}} \sum_\qq \e^{-\ii \RR_j \cdot \qq} \ket{\psi_{\qq, a}} \\
    \ket{w_{j,b}} &= \frac{1}{\sqrt{N}} \sum_\qq \e^{-\ii (\RR_j+\ddelta) \cdot \qq} \ket{\psi_{\qq, b}}
\end{split}
\end{equation}
and, inversely,
\begin{equation}
\begin{split}
    \ket{\psi_{\qq, a}} &= \frac{1}{\sqrt{N}} \sum_j \e^{\ii \RR_j \cdot \qq} \ket{w_{j,a}} \\
    \ket{\psi_{\qq, b}} &= \frac{1}{\sqrt{N}} \sum_j \e^{\ii (\RR_j+\ddelta) \cdot \qq} \ket{w_{j,b}}
\end{split}
\end{equation}
with $\RR_j \in \mathcal{A}$ and therefore $\RR_j+\ddelta \in \mathcal{B}$. The Wannier states $\ket{w_{j,a}}$ and $\ket{w_{j,b}}$ are created by the local creation operators $a_j^\dag$ at position $\RR_j$ and $b_{j}^\dag$ at position $\RR_j+\ddelta$. This basis is convenient as it allows us to exploit our knowledge of the lattice perturbation, as well as the identities following from orthogonality
\begin{equation}
    \begin{split}
        \braket{w_{j,a} | w_{l,a}} &= \delta_{jl} \\
        \braket{w_{j,b} | w_{l,b}} &= \delta_{jl} \\
        \braket{w_{j,a} | w_{l,b}} &= 0.
    \end{split}
\end{equation}
Using the eigenstates of the single-particle Hamiltonian \eqref{eq:vcinab} we can express the states created by $v_\qq^\dag$ and $c_\qq^\dag$ as
\begin{equation}
\begin{split}
    \ket{\chi_{\qq, v}} &= P_\qq^* \ket{\psi_{\qq, a}} + Q_\qq^* \ket{\psi_{\qq, b}} \\
    \ket{\chi_{\qq, c}} &= R_\qq^* \ket{\psi_{\qq, a}} + S_\qq^* \ket{\psi_{\qq, b}}.
\end{split}
\end{equation}

For this derivation, we consider a class of time-periodic lattice perturbations 
\begin{equation}
    W(\rr,t)=W(\rr)\cos(\omega t)
\end{equation}
generated by periodic modulation of the laser intensities $I_i$. The effect of this modulation can be characterized by five real parameters $W_A, W_B, W_1, W_2 \text{ and } W_3$ corresponding to the nonzero matrix elements of $\hat{W}$ in the Wannier basis, where $\hat W$ is the single-particle operator associated with the spatial modulation $W(\rr)$. Here, $W_A$ and $W_B$ denote the modulation amplitudes on the two sublattices, while $W_i$ denotes the modulation-induced hopping matrix element along the nearest-neighbor bond $\ddelta_i\in\mathcal D$, with $\mathcal D=\{\ddelta_1,\ddelta_2,\ddelta_3\}$. Intuition on how the lattice is perturbed to modulate hopping amplitudes along the three bond directions differently is provided by comparing Figs.~\ref{fig:externalpotentialsSM}(a) and (b). We assume that the same modulation pattern is repeated in every unit cell, so that the perturbation preserves lattice translation symmetry.

In the Wannier basis, the corresponding matrix elements are
\begin{equation}
\begin{split}
    \braket{w_{j,a}|\hat W|w_{l,a}} &=W_A\delta_{jl},\\
    \braket{w_{j,b}|\hat W|w_{l,b}} &=W_B\delta_{jl},\\
    \braket{w_{j,a}|\hat W|w_{l,b}} &=W_i \qquad \text{if }(\RR_l+\ddelta)-\RR_j=\ddelta_i\in\mathcal D\\
    &= \sum_{i=1}^3 W_i \delta_{\RR_l+\ddelta - \RR_j, \ddelta_i}.
\end{split}
\label{eq:wannierintoverlap}
\end{equation}

As $\ket{\mathrm{FS}}$ contains $N$ valence-band fermions, whereas $\ket{X_n}$ contains $(N-1)$ valence-band fermions and one conduction-band fermion, only components of $\hat O$ of the form $c^\dagger v$ contribute to Eq.~\eqref{eq:fermisgoldenruleSM}. We are therefore interested in the matrix elements $\bra{\chi_{\kk,c}}\hat W\ket{\chi_{\kk',v}}$, which we now calculate:
\begin{align}
    &\hspace{-0.5cm} \bra{\chi_{\kk, c}} \hat{W} \ket{\chi_{\kk', v}} \notag \\
    &= \left( R_\kk \bra{\psi_{\kk, a}} + S_\kk \bra{\psi_{\kk, b}} \right) \hat{W} \left(P_{\kk'}^* \ket{\psi_{\kk', a}} + Q_{\kk'}^* \ket{\psi_{\kk', b}}\right) \\
    &= \frac{1}{N} \left( R_\kk \sum_j \e^{-\ii \RR_j \cdot \kk} \bra{w_{j,a}} + S_\kk \sum_j \e^{-\ii (\RR_j+\ddelta) \cdot \kk} \bra{w_{j,b}} \right) \hat{W} 
        \left(P_{\kk'}^* \sum_l \e^{\ii \RR_l \cdot \kk'} \ket{w_{l,a}} + Q_{\kk'}^* \sum_l \e^{\ii (\RR_l+\ddelta) \cdot \kk'} \ket{w_{l,b}} \right) \\
    &= \frac{1}{N} \Bigg[ R_\kk P_{\kk'}^* \sum_{j,l} \e^{-\ii \RR_j \cdot \kk}\e^{\ii \RR_l \cdot \kk'} \bra{w_{j,a}} \hat{W} \ket{w_{l,a}} + R_\kk Q_{\kk'}^* \sum_{j,l} \e^{-\ii \RR_j \cdot \kk} \e^{\ii (\RR_l+\ddelta) \cdot \kk'} \bra{w_{j,a}} \hat{W} \ket{w_{l,b}} \notag\\
    &\hspace{1.2cm} + S_\kk P_{\kk'}^* \sum_{j,l} \e^{-\ii (\RR_j+\ddelta) \cdot \kk} \e^{\ii \RR_l \cdot \kk'} \bra{w_{j,b}} \hat{W}  \ket{w_{l,a}} + S_\kk Q_{\kk'}^* \sum_{j,l} \e^{-\ii (\RR_j+\ddelta)\cdot \kk} \e^{\ii (\RR_l+\ddelta) \cdot \kk'} \bra{w_{j,b}} \hat{W} \ket{w_{l,b}}\Bigg].
\end{align}
Inserting \eqref{eq:wannierintoverlap} simplifies this expression further, leading to
\begin{align}
    \bra{\chi_{\kk, c}} \hat{W} \ket{\chi_{\kk', v}}
    &= \frac{1}{N} \Bigg[ R_\kk P_{\kk'}^* \sum_{j,l} \e^{-\ii \RR_j \cdot \kk}\e^{\ii \RR_l \cdot \kk'} W_A \delta_{jl} + R_\kk Q_{\kk'}^* \sum_{j,l} \e^{-\ii \RR_j \cdot \kk} \e^{\ii (\RR_l+\ddelta) \cdot \kk'}  \sum_{i=1}^3 W_i \delta_{\RR_l+\ddelta - \RR_j, \ddelta_i} \\
    &\hspace{1.2cm} + S_\kk P_{\kk'}^* \sum_{j,l} \e^{-\ii (\RR_j+\ddelta) \cdot \kk} \e^{\ii \RR_l \cdot \kk'}  \sum_{i=1}^3 W_i \delta_{\RR_j+\ddelta - \RR_l, \ddelta_i} + S_\kk Q_{\kk'}^* \sum_{j,l} \e^{-\ii (\RR_j+\ddelta)\cdot \kk} \e^{\ii (\RR_l+\ddelta) \cdot \kk'} W_B \delta_{jl}\Bigg] \notag\\
    &= \frac{1}{N} \Bigg[ R_\kk P_{\kk'}^* \sum_{j} \e^{\ii \RR_j \cdot (\kk'-\kk)} W_A + R_\kk Q_{\kk'}^* \sum_{j}\sum_{i=1}^3 \e^{\ii \RR_j \cdot (\kk'-\kk)} \e^{\ii \ddelta_i \cdot \kk'} W_i \notag\\
    &\hspace{1.2cm} + S_\kk P_{\kk'}^* \sum_{j} \sum_{i=1}^3 \e^{\ii \RR_j \cdot (\kk'-\kk)} \e^{-\ii \ddelta_i \cdot \kk}W_i + S_\kk Q_{\kk'}^* \sum_{j} \e^{\ii \RR_j \cdot (\kk'-\kk)} \e^{\ii \ddelta \cdot (\kk'-\kk)} W_B\Bigg].\\
    &= \Bigg[R_\kk P_{\kk}^* W_A + R_\kk Q_{\kk}^* \sum_{i=1}^3 \e^{\ii \ddelta_i \cdot \kk} W_i  + S_\kk P_{\kk}^* \sum_{i=1}^3 \e^{-\ii \ddelta_i \cdot \kk} W_i + S_\kk Q_{\kk}^*  W_B\Bigg] \delta_{\kk, \kk'}.
\end{align}
We define $W_\kk = \sum_{i=1}^3 \e^{\ii \ddelta_i \cdot \kk} W_i$ and use band orthogonality $R_\qq P_\qq^*+S_\qq Q_\qq^*=0$ to obtain the part of the perturbation operator that transfers a fermion from the valence band to the conduction band
\begin{align}
    \hat O &= \sum_\qq \frac12\Bigg[\left(W_A-W_B\right)R_\qq P_\qq^*+R_\qq Q_\qq^*W_\qq+S_\qq P_\qq^*W_\qq^*\Bigg]c_\qq^\dagger v_\qq\\
    &=\sum_\qq T_\qq c_\qq^\dagger v_\qq.
\end{align}
This is the transition operator entering Eq.~\eqref{eq:fermisgoldenruleSM}.

\section{Exciton absorption spectra} \label{sec:spectra}
Here we present numerical results for the absorption spectra calculated using Fermi's golden rule Eq.~(7) with the perturbation operator derived in Sec.~\ref{sec:fermi}. The absorption of an exciton eigenstate $\ket{X_n}$ is calculated using
\begin{equation}
    \bra{X_n}\hat{O} \ket{\text{FS}} = \sum_\qq \alpha_{n\qq}^* T_\qq.
\end{equation}
\begin{figure}[h]
    \centering
    \includegraphics[width=.99\linewidth]{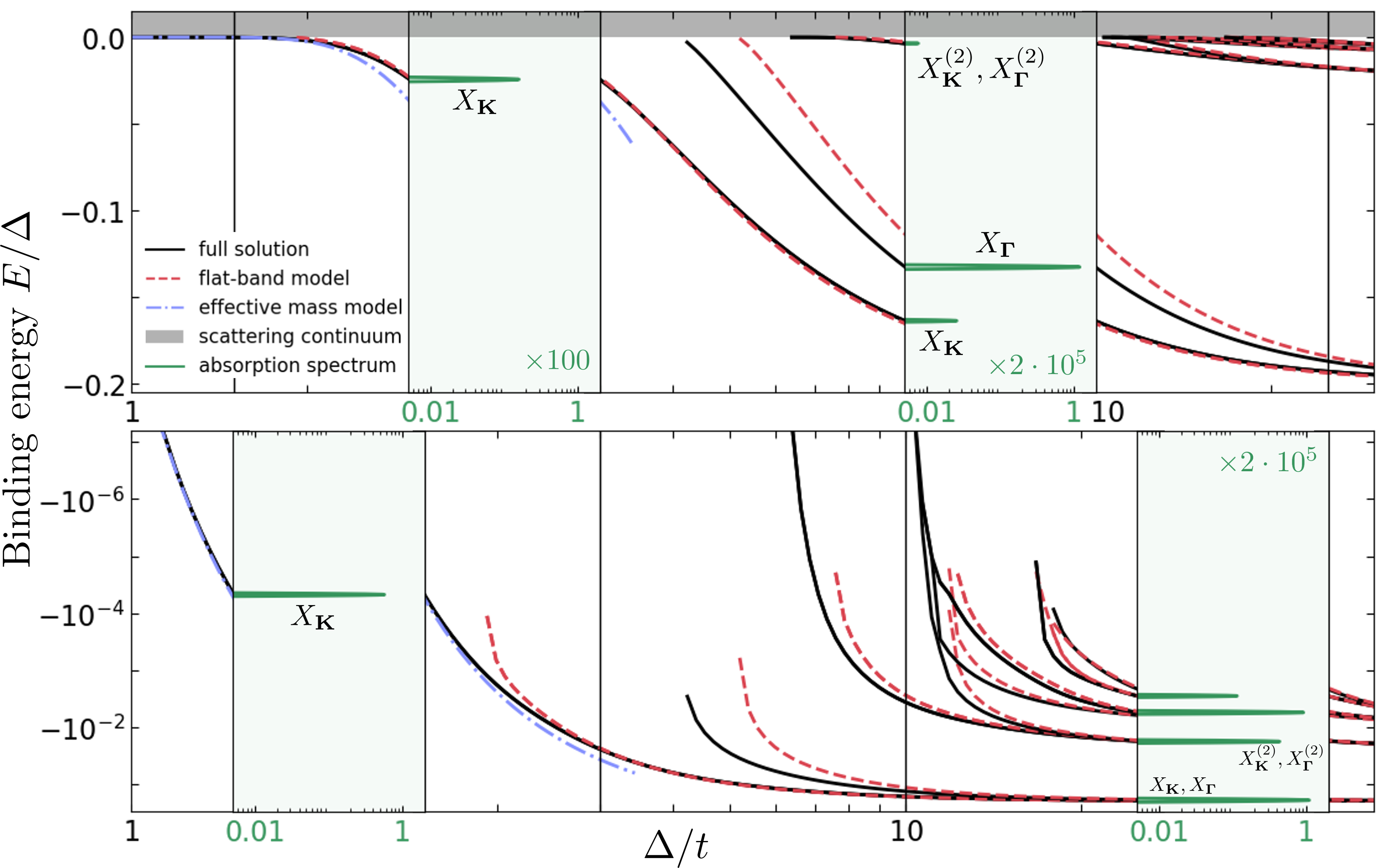}
    \caption{\justifying \textbf{Exciton binding energies and absorption spectra.} Eigenenergies as a function of $\Delta/t$, for fixed $V_{NN}/\Delta=0.2$ over a larger range of $\frac{\Delta}{t}$ than in Fig.~2 in the main text (only the lowest 12 eigenenergies are shown). Bottom: Energies on logarithmic scale. The absorption spectra calculated with $W_A=W_1=1$ and $W_B=W_2=W_3=0$ for $\Delta/t=1.5$, $3$, $10$ and $25$ are shown in green insets. Absorption strength is given in the same arbitrary units over all four panels, i.e. the absorption strength of $X_\KK$ becomes increasingly bright going to larger $\frac{\Delta}{t}$ .}
    \label{fig:excitonenergiesabsorptionSM}
\end{figure}

For the modulation parameters shown in Fig.~\ref{fig:excitonenergiesabsorptionSM}, we find that all eigenstates shown have a nonzero oscillator strength and can therefore be excited. The spectrum shown is robust under variation of $W_A$, $W_B$ and $W_i$ (note that a uniform on-site modulation, $W_A=W_B$ with $W_i=0$, does not couple to any exciton state, while a $C_3$-symmetric bond modulation, $W_1=W_2=W_3$, leaves the states $X_\KK^{(n)}$ dark). Interestingly, we find a hierarchy of ring-localized exciton states in the flat-band regime. This series is expected to continue beyond the results shown here using the same argument of ring-like separation in the relative-position lattice Hilbert space (as given in the main text and shown in Fig.~3(d)).